\documentclass[useAMS,usenatbib]{mn2e}
\usepackage{graphicx}

\title[Star clusters dynamics in a laboratory: electrons in an ultracold plasma]{Star clusters dynamics in a laboratory: electrons in an ultracold plasma}
\author[D. Comparat, T. Vogt, N. Zahzam, M. Mudrich and P. Pillet]{D. Comparat\thanks{E-mail:
Daniel.Comparat@lac.u-psud.fr}, T. Vogt, N. Zahzam, M. Mudrich and P. Pillet\\
Laboratoire Aim\'{e} Cotton, B\^{a}t. 505, Campus  d'Orsay, 91405 Orsay cedex, France
\footnotemark[1]\thanks{Laboratoire Aim{\'{e}} Cotton is associated with Universit{\'{e}} Paris-Sud (website: www.lac.u-psud.fr)}
}
\begin{document}

\pagerange{\pageref{firstpage}--\pageref{lastpage}} \pubyear{2004}

\maketitle

\label{firstpage}

\begin{abstract}

Electrons in a
spherical ultracold quasineutral plasma at temperature in the Kelvin range can be created by laser excitation of an ultra-cold laser cooled atomic cloud. The dynamical behavior of the electrons is similar to the one described by conventional models of stars clusters dynamics. The single mass component, the spherical symmetry and no stars evolution are here accurate assumptions.  The
analog of binary stars formations in the cluster case is three-body recombination in Rydberg atoms in the plasma case with the same Heggie's law: soft binaries get softer and hard binaries get harder. We demonstrate that
the evolution of such an ultracold plasma  is dominated
by Fokker-Planck kinetics equations formally identical to the ones controlling the evolution of a stars cluster. The Virial theorem leads to a link between the plasma temperature and the ions and electrons numbers.
 The Fokker-Planck equation is approximate using gaseous and fluid models. We found that
 the electrons are in a Kramers-Michie-King's type quasi-equilibrium distribution as stars in clusters. Knowing the electron distribution and
 using forced fast electron extraction we are able to determine the plasma temperature knowing the trapping potential depth. 
\end{abstract}

\begin{keywords}
stellar dynamics -- plasmas -- atomic processes -- (stars:) binaries: general
\end{keywords}

\section{Introduction}

One challenge of astrophysics is to understand the dynamics of globular star clusters (for a review see \citet{meylan1997}) because they are test systems for dynamical theoretical models such as
 N-Body modeling, Monte Carlo simulations of Fokker-Planck equations, gas model, scaling models, ...  \citep{binney1987,spitzer1987}. The evolution of globular clusters is dominated by two (or three)-body relaxation, evaporation of stars, tidal truncation and stars evolution. This leads to a very complex evolution and 
the theoretical models are thus simplified with many approximations sometimes far from  reality.
Furthermore
it is not possible to observe the evolution of a given cluster because observation gives only an instantaneous picture.

In this letter we propose to study a real system which 
corresponds to
the most usual assumptions used in conventional models of stars clusters: a single mass component, an almost perfect spherical symmetry and no stars evolution. This system
which can then be 
efficiently compared with theory, namely, an ultra-cold plasma is realized, controlled and studied in a small laboratory. An
ultra-cold plasma can be
 formed by
laser excitation of an ultra-cold ($T\approx 100\ \mu$K) atomic sample and has been first realized by \citealt{kulin1999}.
The physics of ultracold plasmas have strong similarities with the physics of globular stars clusters. Both systems are spherically symmetric, radially limited due to tidal forces for clusters or due to applications of an external magnetic or electric field for plasma. The key point, developed through the whole article, is that both systems are  driven by the same  kinetic equations. Indeed, they are subject to the same inverse square type forces 	if one uses, for the plasma case, a new strong negative ''gravitational constant'' $G'$    defined by:
\begin{equation}
 G'=-\frac{q_e^2/m_e^2}{4 \pi \varepsilon_0} \approx - 2.78\times 10^{32} \rm m^3 .kg^{-1}.s^{-2}\label{replacement} 
	\end{equation}
Analog of binary stars are excited Rydberg atoms and three body recombinations play the same role in both systems. For instance
in a cluster the energy source, in post-collapse evolution,  is provided by binaries formation. Similarly in expanding plasma an heating  is provided by three body recombination and Rydberg atoms formation.
However, some aspects are not perfectly matched in both systems. For instance a globular cluster orbits around its host galaxy and is therefore submitted to centrifugal and tidal forces. On the contrary,
the plasma is not orbiting. Nevertheless, external field (electric or magnetic) might mimic such forces.
In fact the main goal of this article is to demonstrate that, in ultra-cold plasma, the electrons dynamics 
 confined by the ionic potential
  is almost identical to equal mass non evolving stars dynamics in a globular cluster.
 
 The analogy has already been used in $N$-body simulation by \citealt{kuzmin2002a,kuzmin2002b} who simulated the plasma behavior using a modified Aarseth's code usually devoted to cluster studies. In fact the modifications were severe (S. V. Kuzmin personal communication) and 
it is not possible to simply use globular clusters N-Body system equations to fully study ultra-cold plasma system using only physical constant replacement. 
 It is nevertheless possible to precisely study electrons (mass $m_e$, charge $q_e<0$, spatial density $n_e$) behavior within the ionic external potential (ion mass $m_i$, ion charge $q_i=-q_e>0$, spatial density $n_i$) using a true analogy with
a two mass component globular cluster. 
\citealt{vanhaecke2004} were the first to notice a formal analogy and have developed a close analogy using a lowered Maxwellian King's type distribution for the electrons in the plasma. Similar distribution has then been used by \citet{pohl2004d,pohl2004c}. 
 This letter is devoted to an extension of the analogy. The article has been written to be accessible to the ultra-cold plasma community as well as the stellar dynamics community. Lots of ideas, equations or models are not fully solved in this paper. Nevertheless we have though that it is fair and interesting to present preliminary equations in order to stimulate further works.
The outline of the paper is as follows. 
In section \ref{analogy_scale} we  discuss the analogy between the plasma and the cluster system through scaling relation in mass, length and times units. In section \ref{plasma}, we present the basics of ultra-cold plasma physics and the role of the three body mechanism. In section
\ref{FP} we show that, the Fokker-Planck equation is exactly identical for plasmas and clusters system.
In section \ref{gas_model} we briefly study the evolution of 
the quasi-equilibrium distribution using gaseous and fluid model for ions and electrons.
Section \ref{sol_FP} is devoted to the orbit average Fokker-Planck equation. We discuss some
 physical consequence as electron evaporation rate. We detail also some approximate solution  as  a King's type equilibrium distribution
for electrons in the plasma in analogy with globular star cluster dynamics. 
In section \ref{Temperature}
we detail some experimental results. For instance we experimentally extract electrons trapped in an ionic cloud by an external electric field
 to  estimate the electron temperature of the plasma.
 We finally summarize our results and their implications for both astrophysics and plasma physics.

\section{Dynamical analogy through scale units}
\label{analogy_scale}

\begin{table*}
 \centering
 \begin{minipage}{140mm}
  \caption{Typical units value for globular system and for ultra-cold plasma.}
$$\begin{array}{|c|c|c|} \hline
 				& \rm Stars\ cluster & \rm Electrons\ in\  Plasma   \\
\hline \rmn{Number\ of\ particules}   & 5 \times 10^5 &  10^6   \\
\hline \rm System\ Mass  & 5 \times  10^5 \, M_\odot \approx  10^{36} \, \rm kg &  10^6 \, m_e \approx 9 \times 10^{-25} \, \rm kg  \\
\hline \rm Size\ (Core \ radius)  & 1 \, {\rm pc}\approx 3\times10^{16} \ \rm m& 200 \, \mu \rm m = 2 \times  10^{-4} \,  \rm m   \\
\hline \rm Peak\ density  & 8\times 10^3 M_\odot {\rm pc}^{-3}  &   10^{10} {\rm cm}^{-3}  \\
\hline\rm Crossing\ time & 10 \, \rm Myrs \approx 3 \times 10^{15}\, \rm s & 100 \, \rm ns =  10^{-7}\, \rm s  \\
\hline {\rm Velocity\ dispersion\ } \sigma_v & 7 \, \rm km/s  &  50 \, \rm km/s   \\
\hline \rm Relaxation\ time  & 10 \, \rm Gyrs  & 10\, \rm ns  \\
\hline \rm Coulomb\ logarithm: \  \ln \Lambda  & 10  & 4 \\
\hline
\end{array}$$
\label{scaling_table}
\end{minipage}
\end{table*}

In table \ref{scaling_table} we give the typical parameters  of both systems. It is then obvious that the natural units (mass, size, crossing time) for both system are severals tens orders of magnitudes different. But, if we use the units $m_p=m_e,r_p=200\ \mu$m, $t_p=100\,$ns for plasma and  $M_c=M_\odot,R_c=1\,$pc, $T_c=10\,$Myrs for clusters, the systems looks quite similar. 
 Indeed the $k^{\rm th}$ electron evolution in the plasma is described by the Newton's law, with dimensionless scaled notations $\tilde r= r/r_p$, $\tilde  t =t/t_p$ and $\tilde m_e =m_e/m_p$:
\begin{eqnarray}
		\tilde m_e 
		\frac{d^2 \vec{\tilde r}_k}{d \tilde t^2}
	& \approx & - \frac{q_e^2}{4 \pi \varepsilon_0} 
		\frac{t_p^2}{m_p r_p^3} \Big[ \sum_{k_1 \neq k} \frac{\vec{ \tilde r}_{k k_1} }{\tilde r_{k k_1}^3}  +  \label{eq_plasma} \\
	& &	\frac{q_i}{q_e} n_p r_p^3 \frac{\partial}{\partial \vec{ \tilde r}_k}
		 \left(  \int  \frac{\tilde n_i (\vec{ \tilde r})}{\left\| \vec{ \tilde r}-\vec{ \tilde r}_k \right\|} d^3 \vec{ \tilde r} \right) \Big]
		 \nonumber
\end{eqnarray}
where $n_p \tilde n_i = n_i$ is the ion density.

By comparison the full dynamics for an $M_1$ mass star in a two component cluster (stars mass $M_1$ and $M_2$ only) in uppercase dimensionless units $\tilde R= R/R_c$, $\tilde  T =t/T_c$ and $\tilde M_1 =M_1/M_c$ is given by:
\begin{eqnarray}
\tilde M_1 \frac{ d^2 \vec{ \tilde R}_k}{d \tilde T^2} &  \approx  & G M_1^2 		\frac{T_c^2}{M_c R_c^3}\Big[  \sum_{k_1\neq k} \frac{\vec{ \tilde R}_{k k_1}}{\tilde R_{k k_1}^3}  + 
\label{eq_gcluster} \\
& &
\frac{ M_2}{M_1} N_c R_c^3 \frac{\partial}{\partial \vec{ \tilde R}_k}
		\left(  \int \frac{\tilde N_2 (\vec{ \tilde R})}{\left\| \vec{ \tilde R}-\vec{ \tilde R_{k}}\right\|}  d^3 \vec{ \tilde R}  \right) \Big] \nonumber
\end{eqnarray}
 providing $\tilde r =\tilde R$, $\tilde m = \tilde M$ and $\tilde t=\tilde T$ the equation are exactly identical with similar coupling constant: $G' m_e^2
		\frac{t_p^2}{m_p r_p^3}$ and $  G M_1^2 		\frac{T_c^2}{M_c R_c^3} $ on the order of unity (see  equation (\ref{replacement}) and values in table \ref{scaling_table}).

However it is not possible to simply use globular clusters N-Body code to study ultra-cold plasma system. This is
due to the poorly behavior of the second type of particles: ions or $M_2$. 
A first solution is to
modify an existing code devoted to cluster studies as done by  \citealt{kuzmin2002a,kuzmin2002b}.  In a similar manner \citealt{pohl2004c} have used a molecular dynamics simulation with ionic correlations based on the treecode originally designed for astrophysics problems by J. E. Barnes.
Another strategy is to use a Monte-Carlo method. Detail of one Monte Carlo method used for ultra-cold plasma with Three-Body Recombination (TBR) process and radiative atomic lifetime is given by \citealt{robicheaux2003}. Analogy with cluster code as discussed in the MODEST group (MOdeling DEnse STellar systems, http://www.manybody.org/modest/) is obvious and this relation would be probably fruitful for further studies.

\section{Ultracold plasma}
\label{plasma}

\subsection{Ultracold plasma formation}

\begin{figure}
\resizebox{0.45\textwidth}{0.28\textwidth}{
\includegraphics*[0mm,88mm][157mm,204mm]{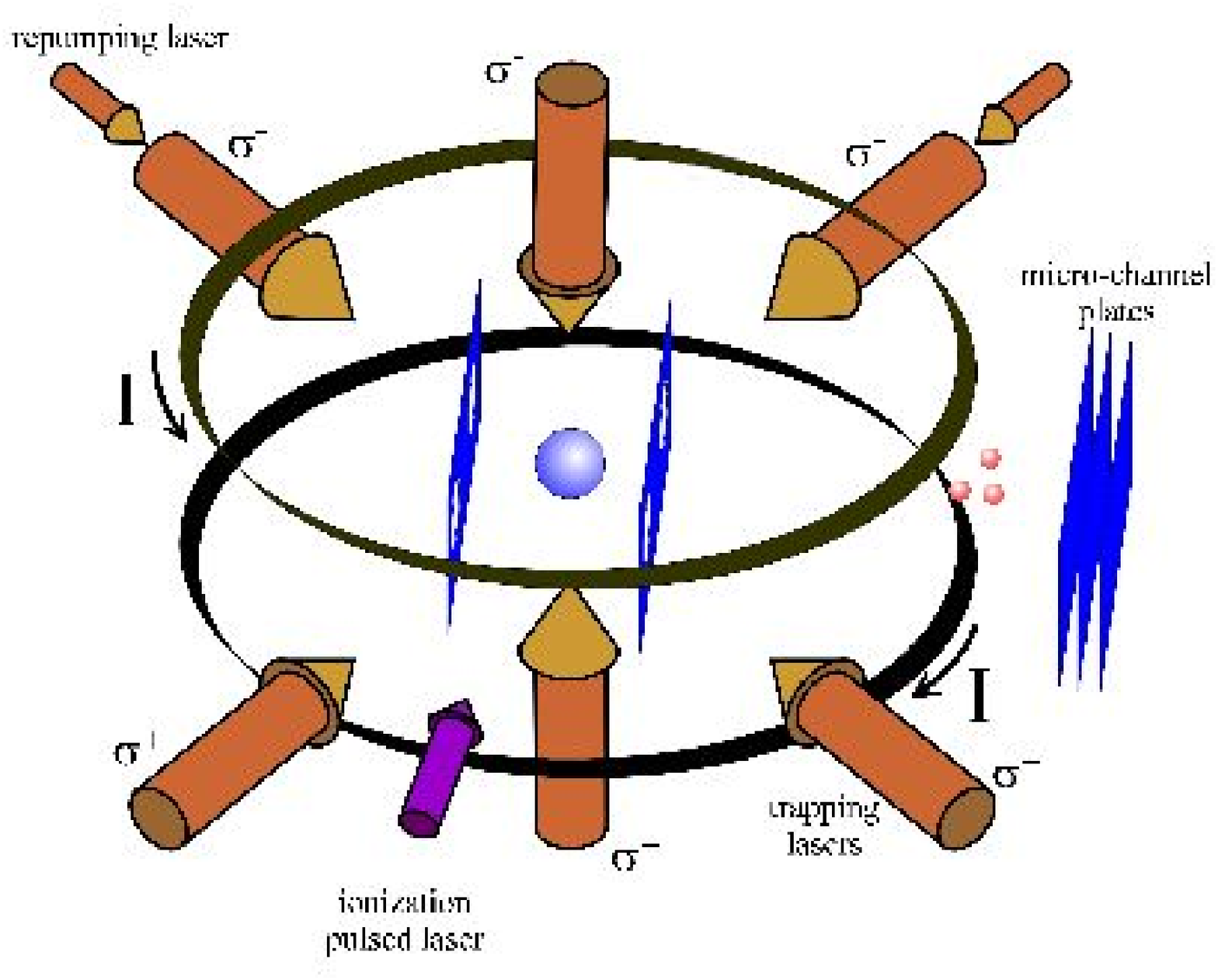}  } 
\begin{center}
\caption{Schematics of a typical experimental setup (typical size $10\,$cm) used to produced an ultra-cold plasma. The whole chamber is under vacuum only filled by the atomic vapor.
The coil (radius of few centimeters), the trapping and repumping lasers are used to laser cooled the atoms. The pulsed dye laser excites the atoms into the ionization continuum or into Rydberg states. The electrons and (or) ions can be extracted by applying electric field on the grid surrounding the plasma. They are then collected by the charged particle detectors MicroChannel Plate (MCP).
}
\label{fig:trap}
\end{center}
\end{figure}

The ultracold neutral plasma physics experimentally began at NIST in 1999 with laser photoionization of a laser-cooled (in the microkelvin or millikelvin range) sample. Since, several groups have done similar work with trapped metastable xenon, cesium or rubidium atoms (for a review see \citep{gallagher2003,killian2003}).  Recently strontium atoms have been also used by \citealt{simien2004}. This is a promising experiment because the strontium ion can be further laser cooled. This might be a way to controlled the ion motions. A typical experimental setup is presented in figure \ref{fig:trap}.
Due to the small electron over ion mass ratio, the electrons have an initial kinetic
energy $k_e^\gamma \equiv 3 k_B  T_e^\gamma /2$ ($k_B$ is the Boltzmann constant) which is, at first glance, believed to be approximately equal to the difference between the photon energy and the ionization threshold. $ T_e^\gamma$ is the photoionisation
 electronic
temperature. 
 Parameters can  easily be experimentally tuned in the following range:
$N_e < 10^8$ electrons at temperature $  T_e^\gamma < 1000\ $K embedded in $N_i< 10^8$ ions (initial temperature $ T_i\approx 1\ $mK).
As indicated in figure \ref{fig:trap} the spherical gaussian atomic sample is ionized by a gaussian laser with an intensity profile given by $I =I_0 e^{-2 (y^2+z^2)^2/w^2}$. This leads to a cylindrical initial plasma shape with $\sigma$ gaussian radius along the longitudinal ($x$) laser propagation axe, $\sigma(t=0)=\sigma_0\approx 250\ \mu$m,  and with $\sqrt{\frac{\sigma^2 \omega^2}{4\sigma^2+\omega^2}}$ gaussian radius along the radial ($y,z$) axes. Experiments are sometimes done with $w\approx \sigma$ so the spherical symmetry is not at all perfect. In the following we will assume $\omega\gg \sigma$ to restore the gaussian spherical symmetry. 
We then have $n_i(r,t)=n_i^0
e^{-r^2/(2 \sigma^2(t))}$, with density $n_i^0<10^{11}\ $cm$^{-3}$. 
 With typical initial value $ T_e = 50\ $K,  $n_i^0 \approx n_e^0 =  10^9\ $cm$^{-3}$, the
 plasma parameters are the following: 
 ion (or electrons) number $N_i=N_e \approx 250\ 000$,
 Debye screening length $\lambda_D = \sqrt{\varepsilon_0 k_B T_e/(q_e^2 n_e^0)} \approx  15\ \mu$m,
 the Wigner-Seitz radius (interparticule spacing) $a_{WS} = ( 4 \pi n_i^0 /3)^{-1/3} \approx 6 \ \mu$m,
  Landau length $r_L = q_e^2/(4 \pi \varepsilon_0  k_B T_e) \approx 0.3\ \mu$m and
 the Coulomb logarithm $\ln \Lambda =\ln (2 \lambda_D/r_L) = \ln (8 \pi q_e^{-3} (\varepsilon_0 k_B T_e)^{3/2} (n_e^0)^{-1/2}) \approx 4$.
 There is several definitions for the electron relaxation time  with slightly different numerical factor 
(as for cluster \citep{louis1991}). We choose here the
 relaxation times  $t_e$ to be defined by $t_e = 9 v_e^3 / (16 \sqrt{\pi} G'^2 m_e^2 n_e^0 \ln \Lambda)  \approx 15 \ $ns for electron with velocity $v_e$, chosen for the numerical results at its r.m.s. value $\sigma_v=\langle (v_e- \langle v_e \rangle)^2 \rangle^{1/2} \approx 50 \, $km/s. The relaxation time is fast enough to have electrons always in  quasi-equilibrium. On the contrary the ions are almost never thermalized with the electrons during the typical $100\ \mu$s  plasma expansion lifetime.
 In the previous definitions
 the temperature has to be understood as the radially dependent thermodynamic (velocity average) temperature 
 $\frac{1}{2} m_e \sigma_v^2 = \frac{3}{2} k_B T_e(r)$. 
  It is worth to note that the plasma is a kinetic plasma because the 
 thermal energy is higher than the
Coulomb interaction energy ($r_L < a_{WS}$). In fact one major goal of the ultra-cold plasma community is to reach the opposite situation, the so called strong coupled regime, where crystallization or correlation can occur. This regime as already been achieved for non-neutral purely ionic (or purely electronic) plasma in trapped system \citep{dubin1999}. 
  
\begin{figure}
\resizebox{0.45\textwidth}{0.28\textwidth}{
\includegraphics*[1.3cm,0.7cm][25.8cm,18.9cm]{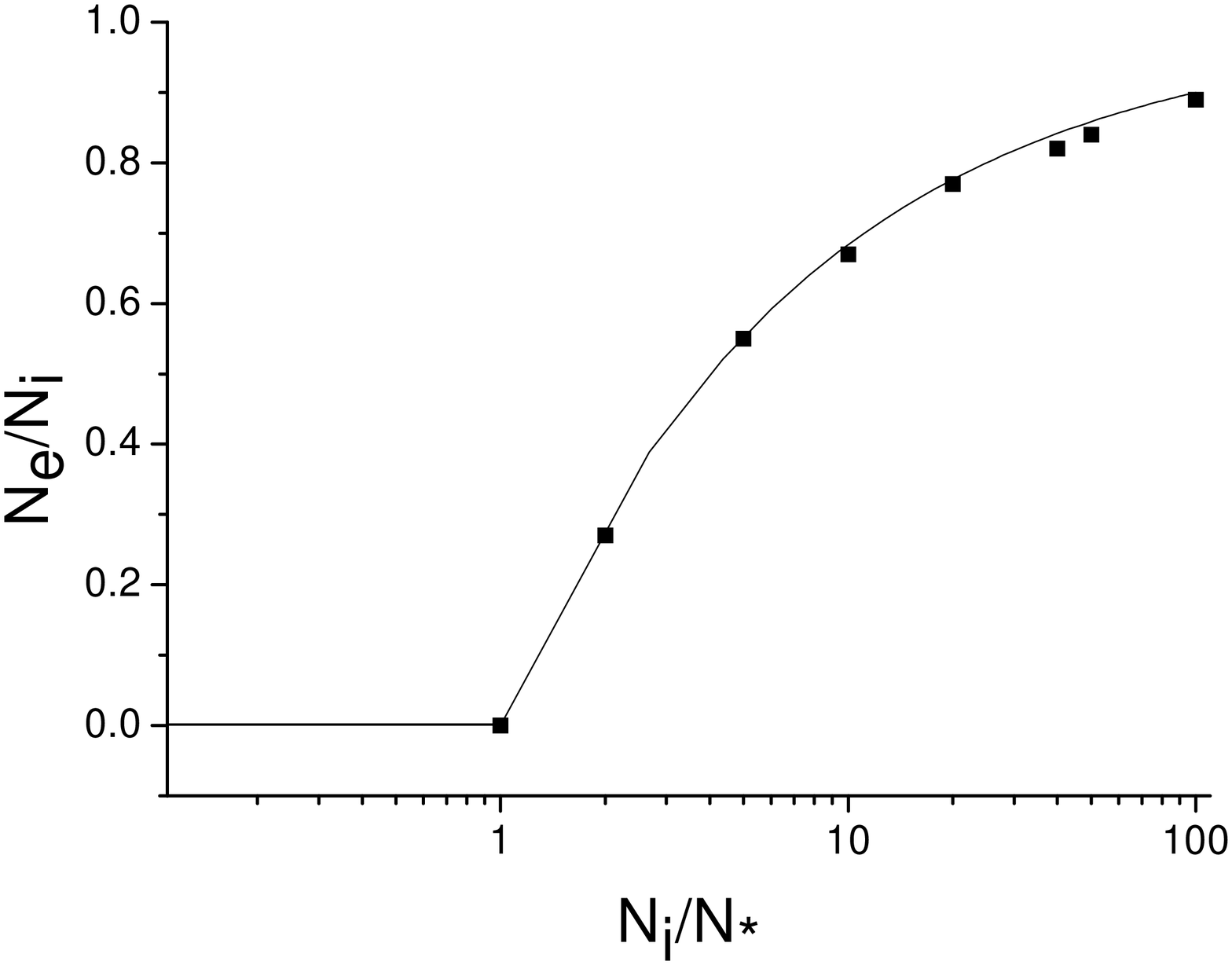} 
} 
\begin{center}
\caption{Experimental results by \citealt{kulin1999}  fitted with solid line by equation (\ref{ratio}).}
\label{fig:nb_ion}
\end{center}
\end{figure}

The physics of ultracold plasma is very rich and only some aspects will be discussed here. Right after the plasma creation, electrons which move faster than ions leave the sample within nanosecond time scale \citep{tkachev2001a}. This very complex stage has not yet been experimentally studied but is related to the Langmuir paradox which is merely the ''violent relaxation'' process in astrophysics \citep{chavanis2002}. This probably leads to almost maxwellian distribution for electrons in a  sub-nanoseconds time scale  which is the inverse of the electron
Langmuir angular frequency
$\omega_{\rm L} \approx \sqrt{4 \pi q_e^2  n_e / (\varepsilon_0 m_e)}$. Similar collisionless process occurs for ions, created in a spatially disordered state, but at hundreds of nanoseconds time scale given by the
inverse of the 
Einstein angular frequency
$\omega_{\rm E}  \approx \sqrt{4 \pi q_e^2  n_i / (\varepsilon_0 m_i)}$ \citep{pohl2004,pohl2004c,simien2004}. The laser ionization create a plasma from where
electrons escape, creating a net positive space charge. When the potential depth becomes equal to the electron kinetic energy, occurring for an ion number equal to,
\begin{equation}
N_* = k_e^\gamma \sigma \frac{ 4\pi \epsilon_0 }{q_e^2} \sqrt{\frac{\pi}{2}} \approx 1500 
\label{N_star}
\end{equation}
the space charge form a trapping potential for the new formed electrons.
  In fact the experimental results by \citealt{kulin1999}, 
  taken for $T_e^{\gamma}$ ranging from 4K to 800K, 
  leads to a more complex result for the final population distribution. We found that \citealt{kulin1999} results can be well fitted (see figure \ref{fig:nb_ion}) by the important formula :
 \begin{equation}
N_i - N_e \approx N_* \left( \frac{N_i}{N_*} \right)^{0.5} 
\label{ratio}
\end{equation}
yielding to 
$N_i-N_e \approx 20000 \ll N_i$ which indicate that the plasma is usually quasineutral. We have obtain similar results, but with a slightly bigger numerical factor in formula (\ref{N_star}). 
Equations (\ref{N_star}) and (\ref{ratio}) can be written in a more convenient form:
\begin{equation}
T_e^\gamma({\rm K}) =  \frac{8.9}{\sigma(\mu{\rm m})}\frac{(N_i-N_e)^2}{N_i}
\label{temp_gamma}
\end{equation}

The electronic pressure finally leads to expansion of the gaussian plasma. \citealt{kulin2000} have experimentally studied this expansion.
They monitored evolution of the mean electron density $\bar n_e$ using electron ejection induced by forced electron plasma oscillations (angular frequency $\omega_{\rm pl} = \sqrt{q_e^2 \bar n_e / (\varepsilon_0 m_e)}$) created by an external Radio-Frequency (RF) electric field \citep{bergeson2003}. The time dependence of the plasma size was then analyzed by Monte Carlo Method by \citealt{robicheaux2002} and may be represented by a typical plasma expansion time $t_{\rm PE}=\sigma_0/v_0$ and an expansion given by:
  \begin{equation}
  \sigma^2(t)\approx \sigma^2_0 + v_0^2 t^2 \ , \  v_0^2 = \frac{k_B T_e^\gamma }{m_i} \label{expansion}
  \end{equation}

In fact after typically few microseconds an ionic density spike appears \citep{robicheaux2003} and formula (\ref{expansion}) is no more valid but seems to be restored for much longer time \citep{pohl2004c}. 
One major experimental result \citep{kulin2000} concerns the fact that even for laser ionization at threshold, where a near zero electron velocity is naively expected, the velocity $v_0$  is no more given by formula (\ref{expansion}) but is experimentally found to be always greater than $40\ $m/s and greater than $  \sqrt{ k_e^\gamma / m_i}$ \citep{kulin2000}. One gets the fundamental result that the electron temperature $T_e$ is higher than $5\,$K independently of the experimental initial parameters. The plasma is therefore always in a non coupled regime where $r_L<a_{WS}<\lambda_D$.
 This, can be due to instantaneous initial 
increase electron kinetic energy while moving toward ions \citep{kuzmin2002a}
or due to
continuum lowering \citep{hahn2002,mazevet2002} which reflects the fact the 
energy of the isolated atoms is shifted by long-range Coulomb interaction with neighbors when embedded in a plasma.
Another probable explanation results from
 huge three body recombinaison  (TBR) rate.
 In such collision between two electrons and one ion,
 (Rydberg) atoms are formed in the plasma when one electron recombine with the ion. 
  \citealt{killian2001}  have indeed observed in ultra-cold plasma  this TBR. Rydberg atoms formation are the analog of
binary system formation through three body encounters in globular cluster, this is a large energy source for the free remaining electron which strongly heats the sample \citep{robicheaux2002}. 

\subsection{Three body recombination processes, binary system and Rydberg atoms}
\label{TBR}

The reverse process namely spontaneous evolution of an ultracold excited (Rydberg) gas to an ultra-cold plasma  have been
demonstrated by our group and by a group at University of Virginia simultaneously in year 2000 \citep{robinson2000}. 
 The Rydberg ionization process probably starts with blackbody photoionization or high energetic collisions with hot surrounding atoms. The so formed initial electrons leave, almost instantaneously, the
interaction region. One attempted analogy with cluster might be dissociation of primary binary system due to external supernovae explosion but with much higher probability rate.
 A second phase occurs when the positive ion potential is deep
enough to trap subsequent electrons, which then collide with Rydberg atoms creating more electrons in an avalanche ionization
process~\citep{robicheaux2003,pohl2003}. This is perfectly identical to binary collision in stars clusters when collision with a third star leads to  destruction of the binary system. 
However, other relevant processes have been proposed whose effects need to be
investigated, such as  autoionization of Rydberg atom pairs \citep{hahn2000} similarly to destruction of binary system through binary-binary collisions in globular clusters.
\citealt{vanhaecke2004} have reported the total ionization of a cold Rydberg atomic sample embedded in an almost neutral ultracold plasma. This experiment is the analog of primordial binary system present initially inside a globular cluster. As in the astrophysical point of view this has strong influence on the energetic story of the sample. Rydberg atoms have finite lifetime (typically tens of microsecond) induced by  photon spontaneous emission or by blackbody photon absorption; these processes can be seen as the analog of stellar evolution.
Some other works have
shown that high-angular-momentum  (circular electron orbit)  Rydberg atoms, with order of magnitude longer lifetime (several millisecond), are created by eccentricities change through collision with electrons in an ultra-cold plasma \citep{dutta2001,walz2004}. This kind of process has been also mentioned  in case of ZEKE (Zero Kinetic Energy) photoelectron spectroscopy  \citep{hua2001}. The high-angular-momentum Rydberg atoms are the analog of circular binaries in cluster environments which is still an open study \citep{giersz2004}.  \citealt{walz2004} also note that there is an unexpected large population of deeply bound (hard) Rydberg atoms, this is probably due to Penning ionization where two binaries collide leading to the disruption of one of them and increasing of the binding energy of the second. This Penning ionization is the exact analog of the disruptive collision in binary-binary scattering in the cluster case \citep{spitzer1987,hut1992}. Indeed,
the collisions during binary-binary and binary-single interactions still need a lot of investigations \citep{fregeau2004}. Another interesting study is relative to the binding energy distribution. Starting from a pure plasma the  binding energy distribution, resulting of the TBR process,  is modified during the plasma expansion \citep{robicheaux2002}.
The center of mass Rydberg velocity is the one of the ion at the time the recombination occurs and should therefore be linked with the Rydberg binding energy: the slower atoms tend to have larger binding energies \citep{robicheaux2003}. This is also linked to the so-called excitation freezing \citep{tkachev2001} and needs to be experimentally studied.
There is however differences between Rydberg atoms and binary stellar systems \citep{hut1983} (see also the nice review on binaries in globular stars clusters by \citealt{hut1992}). Indeed, Rydberg are dipolar but neutral atoms. Therefore the Rydberg binary system interact with neighbors only through charge-dipole or dipole-dipole interaction scaling respectively as $a_{WS}^{-2}$ and $a_{WS}^{-3}$ but in the cluster case binary system interact through Coulomb interaction scaling $a_{WS}^{-1}$. Furthermore,
 the dissipation of the relative kinetic energy of two strongly interacting stars through tidal effect which lead to the formation of tightly bound binaries (so called tidal binaries) has no simple analog in the plasma case. 
The goal of the article is to give an overview of the analogy between both systems, but it is beyond its scope to detail all the process involving binaries. We will mainly focus on results which are of interest for the free electrons distribution.

When a Rydberg gas is formed, there is competition between
the deexcitation rate and the excitation rate for Rydberg atoms. 
In fact, very highly excited Rydberg atoms, for which binding energy are less than the single electron plasma kinetic energy, are analog to soft binary stars  and, as in the cluster case, encounters lead to disruption of the system by gradually increasing the binding energy until it becomes positives. 
 This is a similar law than Heggie's law (we will use this terminology even if this law was known in atomic physics much earlier than in cluster physics) : hard binaries get harder and soft binaries get softer \citep{heggie1975}. Indeed,
the probability of excitation for a Rydberg atom equals the 
 probability of deexcitation when its
 binding energy is given by $3.82 k_B T_e$ \citep{mansbach1969,robicheaux2003}. A similar formula but with a bottelneck point nearer to $k_B T_e$ has been given by \citet{tkachev2001}, see also \citet{stevefelt1975} or \citet{vriens1980} for other formulas. This is exactly the same behavior than the one noticed in stars cluster physics. This 
  process have been studied experimentally  \citep{killian2001,li2004,walz2004}.

As previously mentioned, in both systems, the binaries formation comes from three body recombinaison  (TBR). For cluster, the rate
$\Gamma_{\rm TBR}$  is \citep{binney1987} (8-7) and \citep{spitzer1987} (6-37)
$$
\Gamma_{\rm TBR} \approx \frac{n^2 G^5 M^5}{\sigma_v^9} .
$$
where $\sigma_v^2=k_B T_e/m_e$ is the one dimensional velocity dispersion.
And for electrons plus ions we have \citep{mansbach1969,tkachev2001}
\begin{equation}
\frac{1}{n_e} \left( \frac{d n_e}{d t} \right)_{\rm TBR} = - \Gamma_{\rm TBR} \approx \frac{ n_e n_i G'^5 m_e^5}{ (k_B T_e/m_e)^{9/2}} \approx - 100 \, {\rm s}^{-1}
\label{TBR_rate}
\end{equation}
In both formula we have omitted numerical factor of the order of unity. The analogy is obvious but
the $T_e^{-9/2}$ behavior leads to a main difference between the cluster case and the ultra-cold plasma case. For an ultracold plasma the TBR rate is huge but it is almost negligible for clusters.
We mention that the analogy could be pursuit further on. Indeed, formula given by \citep{sigurdson1993} (see also \citet{hut1983,spitzer1987}) which concern the energy change between stars is similar to the rate function \citep{mansbach1969} (III-12) between atomic energy states ${\cal E}_i = k_B T_e \epsilon_i$ to energy state ${\cal E}_f = k_B T_e \epsilon_f$ given by:
\begin{eqnarray}
k(\epsilon_i,\epsilon_f) &=& k_0 (- \epsilon_f)^{-4.83} (-\epsilon_i)^{2.5}, \ \epsilon_i>\epsilon_f \label{eqn_rate_MC} \\
k(\epsilon_i,\epsilon_f) &=& k_0 (- \epsilon_i)^{-2.33} e^{-(\epsilon_i-\epsilon_f)}, \  \epsilon_i<\epsilon_f \nonumber
\end{eqnarray}
where $k_0 = 11 \left( \frac{q_e^2}{k_B T_e} \right)^2 \left( \frac{k_B T_e}{m} \right)^{1/2} $ and the zero of energy is taken as the ionization threshold.

In the cluster case, the energy heating rate due to binaries formation is 
$$
\dot {\cal E} \approx 100 n^2 G^5 m^6 \sigma_v^{-7} \approx 100 \Gamma_{\rm TBR} m \sigma_v^2
$$
the factor hundred (which is a very approximate one) comes from the fact that every binary liberates in the cluster an energy of few hundreds times $k_B T$ through encounters with other stars before being ejected out of the cluster by a very energetic reaction \citep{cohn1989}. This non-negligible flux of escaping binaries is a 
 well known effect in cluster case \citep{meylan1997}. But, flux of escaping binaries is probably negligible in the plasma case, because the binding energy of the binary increases mainly due to radiative lifetime and not much due to collisions. The bottleneck occurs when its energy becomes lower than 
 $
 \epsilon^* \approx k_B \times 500\,{\rm K}\times \left( \frac{T_e}{1\,{\rm K}} \right)^{-2/9}\times \left( \frac{n_e}{10^9\,{\rm cm}^{-3}} \right)^{1/9}
 $ \citep{tkachev2001}.
Therefore
in the plasma case, we do not have such simple results for $\dot {\cal E} $ because the radiative lifetime combined with the expansion play a complex role, but we could estimate \citep{tkachev2001} (12):
\begin{equation}
\dot {\cal E} \approx 5.4 \left( \frac{n_e^0}{10^9 {\rm cm^{-3}}} \frac{t_{\rm PE}}{3  \mu s} \right)^{-2/9} 
 \Gamma_{\rm TBR} k_B T_e
 \label{heat_rate}
\end{equation}
The analogy holds because the process evolve as $\dot {\cal E} \propto  \Gamma_{\rm TBR} k_B T$ in both cases.

Finally, it might be useful, ofr further studied, to note that if the time dependence Rydberg binding energy distribution is complex (see \citet{pohl2004c,robicheaux2003}), we found that the results can be well fitted by a binding energy ${\cal E}_{\rm Ryd}>0$ distribution of the Rydberg states proportional to
\begin{equation}
e^{-{\cal E}_{\rm Ryd}/(k_B T_{\rm Ryd})} \left( \frac{{\cal E}_{\rm Ryd}}{k_B T_{\rm Ryd}} \right)^{-\alpha} \label{Ryd_en_dist}
\end{equation}
 with $\alpha$ and $T_{\rm Ryd}$ are time dependent parameters and $\alpha$ is always smaller than its equilibrium value of $5/2$.



\section{Fokker-Planck equation}
\label{FP}

\subsection{Symmetry consideration}

\citealt{rosenbluth1957} already indicated that the Fokker-Planck equation is formally identical for electrons in plasma or for stars in stellar
 systems. 
We thus expect similar behavior for our system or for  an isolated cluster.
Velocity anisotropy is generated by two-body relaxation in the outer part of a globular system \citep{spitzer1987}. We expect similar results here but,
to simplify the discussion, we will use isotropic symmetry.
Using Lynden-Bell's improved strong Jean's theorem for a spherically symmetric plasma system \citep{binney1987} or assuming ergodicity in our ionic non harmonic trapping potential \citep{surkov1996}, we could assume
further on  that the electronic phase density function $f(\vec r,\vec v,t)$ in the ultra-cold plasma depends only on energy
$  {\cal E} =q_e \phi+ \frac{1}{2} m_e v_e^2 $ where $\phi=\phi_e+\phi_i$ is the sum of the electronic and ionic potential.
To be able to compare with stellar dynamics it is more convenient to use energy per mass notation $E =  {\cal E}/m_e =\Phi + \frac{1}{2} v_e^2$.  The potential energy (per mass units) is
$\Phi = \Phi_i+\Phi_e=q_e(\phi_i+\phi_e)/m_e $.  
 Using the Poisson's equation one gets:
\begin{eqnarray}
	\Delta \Phi & = &\frac{1}{r^2} \frac{\partial}{\partial r} \left( r^2  \frac{\partial \Phi}{\partial r} \right)  = - \frac{q_e}{m_e} \frac{  q_e n_e +  q_i n_i }{\varepsilon_0} = 4 \pi G' (\rho_e+\tilde \rho_i)  \nonumber
 \\
  \frac{\partial \Phi }{\partial r} &= & \frac{G' M^{\rm tot}_r}{r^2} = G' m_e \frac{N_e(r) -  N_i (r)}{r^2} > 0 \label{Eq_Poisson}  \label{sol_Poisson} \\
	 \Phi(r) & = & 	 \Phi_\infty -  4 \pi G' \left( \frac{1}{r} \int_0^r \rho^{\rm tot} (r')  r'^2 d r' + \int_r^\infty \rho^{\rm tot} (r') r' d r' \right) \nonumber
	\end{eqnarray}	
where $\rho_e=m_e n_e$ is the mass density for electrons, $ \rho^{\rm tot}=\rho_e+\tilde \rho_i$ , and $\tilde \rho_i = - m_e n_i$ which is not the ionic mass density $\rho_i = m_i n_i$.
The artificial ''total'' mass is $M^{\rm tot}=(M_e+ \frac{q_i m_e}{q_e m_i}  M_i)=m_e(N_e-N_i)=M_e+\tilde M_i <0$.
 $N_e(r)=M_r/m_e$ (respectively $N_i(r)$) is the number of electrons (respectively ions) inside a sphere of r radius:
 \begin{equation}
 N_i(r) =  \left(  {\rm Erf}\left(\sqrt{\frac{r^2}{2 \sigma^2}}\right)
  - \sqrt{\frac{2 r^2 }{\pi \sigma^2 } } e^{-\frac{r^2}{2 \sigma ^2}} \right) 
   {N_i} \label{Nb_ion}
 \end{equation} 
  $\rm Erf$ is the error function and
  \begin{equation}
 \Phi_i(r) =  G m_e N_i   \frac{ {\rm Erf}\left(\frac{r}{\sqrt{2} \sigma}\right) }{\frac{r}{\sqrt{2} \sigma}}
 \label{phi_gauss} \label{pot_gauss}
 \end{equation} 

The magnetic field (see coils in figure \ref{fig:trap}) is usually turned off during an experiment or its effect is small and can be neglected in first approximation.
		The Fokker-Planck equation for the electron space phase density distribution $f=f_e$, 
	which can be seen as a series in $\ln \Lambda$ and is then no more valid in the strongly coupled case ($\Lambda<1$),
		is \citep{delcroix1994,mitchner1992}
\begin{eqnarray}
\frac{{\rm d} f}{{\rm d} t} & = &  \sum_f \frac{\Gamma_f}{ v^2} 
 \frac{\partial}{\partial v} 
 \Big[ f \frac{m_e}{m_f}
 \int_0^v  f_f  4 \pi v_f^2 d v_f   +  \label{FP_v}   \\
 & & 
\frac{v}{3}  \frac{\partial f}{\partial v}  \left( \frac{1}{v^2} 
\int_0^v  v_f^2 f_f  4 \pi v_f^2 d v_f 
+ v  
 \int_v^\infty \frac{1}{v_f} f_f 4 \pi v_f^2 d v_f
 \right)
\Big] \nonumber
\end{eqnarray}
with
$$
\frac{{\rm d} f}{{\rm d} t}  =   \frac{\partial f}{\partial t} + \vec v \cdot \frac{\partial f}{\partial \vec r}
-  \frac{\partial \Phi}{\partial \vec r}  \cdot \frac{\partial f}{\partial \vec v} 
$$
where the subscript $_f$ indicates the type of field particles (electrons or ions).
$\Gamma_f = 4 \pi G^2 m_f^2 \ln \Lambda$  for clusters becomes
$ \Gamma_f = 4 \pi G'^2 m_e^2 \ln \Lambda = \frac{q_e^2 q_i^2}{4 \pi m_e^2 \varepsilon_0^2}  \ln \Lambda = \Gamma$ in the plasma case which, in this case, is independent of the field particle mass.

If we use the so called thermal bath (i.e. Maxwellian) approximation for the field distribution function $f_f$, one gets:
\begin{equation}
 \frac{{\rm d} f}{{\rm d} t}  = \sum_f
\frac{\Gamma_f n_f }{ v^2} 
 \frac{\partial}{\partial v} 
 \left[ \tilde G( \frac{\sqrt{3} v }{\sqrt{2  \langle v_f^2 \rangle} } ) 
 \left(
 f \frac{m_e}{m_f}+ \frac{ \langle v_f^2 \rangle }{v^2}
\frac{v}{3}  \frac{\partial f}{\partial v}   \right)
\right] \label{thermal_bath}
\end{equation}
Where
$\tilde G (x) = {\rm Erf} (x) - x \frac{d {\rm Erf}}{d x} (x)  $
 tends to $1$ when $x$ tends to infinity and is proportional to $x^3$ for small x values.
Using relations 
$T_i \ll T_e$ (or $\langle v_i^2 \rangle \ll \langle v_e^2 \rangle $) and $
n_i \frac{m_e}{m_i} \ll n_e
$  we found  that in equation (\ref{thermal_bath}) the collisional ionic terms
are negligible,
 in the field particles collisional part, compared to the electronical ones. with Fokker-Planck  of pure electrons. Collisional ions effect can be summarized mainly as 
$n_i \frac{m_e}{m_i}$ added to $n_e$ and a $\langle v_i^2 \rangle  $ terms added to $v^2=v_e^2$ in the electronical part. 
The arguments hold also for non maxwellian distributions and the key results is that we can always safety neglect
ions in the collisional part. Finally the (Landau)-Poisson-Fokker-Planck equation for electrons
		is then \citep{delcroix1994,mitchner1992}
\begin{eqnarray}
 \lefteqn{\frac{\partial f}{\partial t} + \vec v \cdot \frac{\partial f}{\partial \vec r}
-  \frac{\partial \Phi}{\partial \vec r}  \cdot \frac{\partial f}{\partial \vec v}  =   \frac{4 \pi \Gamma}{ v^2} 
 \frac{\partial}{\partial v} 
 \Big[ f \int_0^v  f'  v'^2 d v'
   + } \nonumber \\
   & & 
\frac{v}{3}  \frac{\partial f}{\partial v}  \left( 
 \int_0^v  f' \frac{ v'^4}{v^2} d v' + \int_v^\infty  f'  v v' d v' \right)
\Big] \label{FP_tot}
\end{eqnarray}
 the prime indicates that the quantity depends on the encounter field velocity $v'$, not on $v$. This equation
contains only electrons  and is thus  completely identical (using equation (\ref{replacement})) to the Fokker-Planck equation for a single mass stars system.

\subsection{Virial theorem}
There is fundamental uncertainties in the value of 
  $\ln \Lambda$. Formula $\ln \Lambda = \ln ( 8 \pi q_e^{-3} (\varepsilon_0 k_B T_e)^{3/2} (n_e^0)^{-1/2} )$ 
is
only correct for homogeneous systems where $n_e(r)=n_e^0$. 
For a star system an usual approximation is \citep{spitzer1987} (2-14), $ \Lambda \approx 2\frac{1.5 r_h}{ \tilde r_L} $ 
 where $r_h$ the  radius containing half mass of the system and $\tilde r_L = \frac{ G  m^2 }{ k_B \bar T }$. Similarly for plasma we define a (global) temperature $\bar T_e$ \citep{binney1987} by:
$$
2 K_e  \equiv  3 k_B \bar T_e N_e  \equiv 2 \int d^3 \vec r d^3 \vec v \frac{1}{2}
  m_e \vec v^2 f(\vec r,\vec v,t) 
$$
where $K_e$ is the total electron kinetic energy. By analogy,
in the plasma case, we can write
$\Lambda =  2 \frac{ \lambda_D}{r_L} $ with 
  $r_L = - G' m_e^2  /  k_B \bar T_e $.
 The equations are then perfectly identical  for both systems except the presence of the Debye screening distance in the Coulomb logarithm for plasma but the radius limit distance for cluster.
One usual average value, used in single mass globular clusters,  consists to evaluate $r_h$ using the 
potential energy 
\begin{equation}
W^{\rm tot} = \int w^{\rm tot} \equiv   \frac{1}{2} \int_0^\infty  4 \pi r^2 \rho^{\rm tot} (r) \Phi (r) d r \approx - 0.4 \frac{G M^2}{r_h}
\label{pot_tot}
\end{equation}
where the constant $0.4$ is a reasonable approximation for most systems ($0.44$ for gaussian density distribution
where 
 $r_h\approx 1.54\sigma$) \citep{spitzer1987}.
The static Virial theorem $2 K  = - W^{\rm tot}$ leads 
 to  
$\Lambda \approx   0.4 N $
 where $N$ is the number of stars in the cluster.
 In the  plasma case the
  Virial theorem \citep{binney1987} is:
\begin{eqnarray}
 2 K_e
& =  & \int_0^\infty \rho_e(r) r \frac{\partial (\Phi^e+\Phi^i)}{\partial r}  4 \pi r^2 
 d r  \nonumber \\ 
 &= & -2 W_{ee} - W_{ei} 
  \end{eqnarray}
 Where  $W_{ei}=\int_0^\infty \rho_e(r) r \frac{\partial \Phi^i}{\partial \vec r}  4 \pi r^2 
 d r $ can be seen as an external ''potential'' energy (due to the ions) and $W_{ee} = \frac{1}{2} \int \rho_e \Phi_e $.
 A naive extrapolation, based on quasineutrality approximation ($n_e-n_i$ constant), of the cluster case would leads, in the plasma case to:
 \begin{eqnarray}
2 K_e  =  3 k_B \bar T_e N_e &\approx& 0.4 \frac{G' M^{\rm tot} M_e}{r_h} \label{Virial_plasma} \\
 \bar T_e ({\rm K}) 
&\approx& 1.6 (N_i - N_e)/\sigma (\mu{\rm m}) \label{virial_res}
 \end{eqnarray}
  which have to be compared to formulas (\ref{pot_tot}), (\ref{temp_gamma}). 
Finally, using this naive evaluation of the Virial theorem we could extend for the plasma the formula $\Lambda \approx   0.4 N $ valid for clusters in:
\begin{eqnarray}
\Lambda & = & 0.4 (N_i - N_e) \frac{2\lambda_D}{3 r_h} \\
& \approx & 0.1  (N_i - N_e) \sqrt{\frac{N_i - N_e }{N_e}}  \nonumber
\end{eqnarray}
where we have used formula (\ref{virial_res}) and the gaussian approximation $\lambda_D\approx  \sqrt{ (2 \pi \sigma^2)^{3/2} \varepsilon_0 k_B \bar T_e/(q_e^2 N_e)}$ to derive the final formula.
We will see later that electrons are not in a gaussian distribution and that these naive formula have to be corrected. Nevertheless,
they indicate three very important results. Firstly the ions minus electrons number is directly related to the temperature. 
 Secondly, the final temperature $\bar T_e $ is on the same order to the one expected from the laser wavelength $T_e^\gamma$ . 
 Thirdly, the coupled regime ($\Lambda<1$)  can be reached only for almost pure neutral plasma with numerous electrons.
 
\section{Fluid and gas models}
\label{gas_model}

\subsection{Gaseous equations}
 One of the easiest way toward an approximate solution of the electron Fokker-Planck equation (\ref{FP_tot}) is to use the velocity moment equation. Here again, to avoid confusion and to be able to use equations derived for stars dynamics we define
 the mass density distribution $f_m= m_e f $.
 It is simple to use dimensionless equations where $r=r_0 r^*, t=t_0 t^*, \rho(r,t)=\rho_0 \rho^*(r^*,t^*), M_r(r,t)=M_0 M^*(r^*,t^*), \ldots$ or the selfsimilar form where $r=r_c(t) r_*, \rho(r,t) = \rho_c (t) \rho_*(r_*),  \ldots$ \citep{louis1991}.
  We then define
the mass density $\rho = \rho_e = m_e n_e = \langle 0 \rangle  = \int d^3 \vec v f_m(\vec v) = \int_0^{+ \infty} 4 \pi v^2 f_m(v) d v$ related to $M_r=m_e N_e (r)$ the mass contained in a sphere of radius $r$. We also define
the velocity of mass transport 
$u$ by $\rho u = \langle 1 \rangle  = \int  v_r f_m(\vec v) d^3 \vec v$ where $v_r$ is the radial velocity, the kinetic energy density 
$k_E$ by $ 2 k_E = \langle 2 \rangle   = \int_0^{+\infty} 4 \pi v^2 v^2 f_m(v) d v$. This also define the pressure $p$ by $ 3 p + \rho u^2 = 2 k_E $ which is linked to the temperature $T_e$, the one dimensional velocity dispersion $\sigma_v$ (isothermal sound speed)  through $p=n_e k_B T_e = \rho \sigma_v^2 $.

The moments equations are the following:
\begin{itemize}
	\item The mass integration
\begin{equation}
	\frac{\partial M_r}{\partial r} =   4 \pi r^2 \rho  
	\end{equation} 
which can be written as
\begin{equation}
\frac{\partial \ln M^*}{\partial \ln r^*} = {r^*}^2 \rho^* \nonumber
\end{equation}
for the choice $M_0=4 \pi r_0^3 \rho_0$

\item The continuity equation (with $u_0=r_0/t_0$)
\begin{eqnarray}
	0 & = &	\frac{\partial \rho}{\partial t} + {\rm div} (\rho u) = \frac{\partial \rho}{\partial t} + \frac{1}{r^2} \frac{\partial r^2 \rho u}{\partial r} \\
	 0 & = & \frac{\partial \rho^*}{\partial t^*} +	\frac{u^*}{r^*} \frac{\partial \ln \rho^*}{\partial \ln r^*} +
	 \frac{\partial u^*/r^*}{\partial \ln r^*} + 3\frac{u^*}{r^*} \nonumber
\end{eqnarray}
\item The hydrodynamical (Euler) equation
\begin{equation}
0  = 	\frac{D u}{D t} + \frac{1}{\rho}  \frac{\partial p}{\partial r} + \frac{\partial \Phi }{\partial  r}  	\label{hyd_dyn}
\end{equation}
where $\frac{D }{D t} = \frac{\partial}{\partial t} + u \frac{\partial}{\partial r}$ is the co-moving (Lagrangian or convective) derivative  following the mass evolution $\frac{D M_r}{D t}=0$. 

 The third momentum equation is :
 \item The kinetic energy transport equation
\begin{equation}		\frac{\partial  k_E}{\partial t} + \frac{1}{2} {\rm div}  \langle 3 \rangle +    \rho u \frac{\partial \Phi}{\partial r} = 0 \label{ene_cons}
	\end{equation}
	
The fact all these equations do not contain right hand side collisional terms is  a verification that the Fokker-Planck equation conserve the mass and the energy. Therefore there is no difference between the use of the collisionless Boltzmann (also called Vlasov) equation and the use of the collisional Fokker-Planck equation at this stage.

\subsection{Self similar collisionless quasineutral plasma evolution}

The main goal of this paper is electrons evolution, but
similar gaseous equations holds for the ions in the plasma. In the ion
case 
 the pressure is negligible ($T_i \approx 0$) and the system of gaseous equation is closed with
 \begin{equation}
 0 = \frac{D u_i}{D t}  +  \frac{1}{m_i n_i} \frac{\partial p_i}{\partial  r} + \frac{q_i}{m_i} \frac{\partial \phi}{\partial  r} \approx \frac{D u_i}{D t}  +  \frac{q_i}{q_e} \frac{m_e}{m_i} \frac{\partial \Phi}{\partial  r} =0.
 \end{equation}
 The quasineutrality equations: $q_i n_i+q_e n_e \approx 0 $ and $q_i\frac{D u_i}{D t}+q_e\frac{D u_e}{D t}=0$ can be used, with the Euler equations for ions and electrons, to lead to a useful relation between the potential and the pressure :
 \begin{equation}
 \frac{\partial \phi}{\partial   r} \approx \frac{
 \frac{q_i}{m_i } \frac{\partial p_i}{\partial  r} +\frac{q_e}{m_e } \frac{\partial p_e}{\partial  r}
 }{
 \frac{q_e^2 n_e}{m_e} + \frac{q_i^2 n_i}{m_i} 
 }
 \label{equ_self_sim}
 \end{equation}
 
 Assuming Maxwell-Boltzmann electron distribution we found
$$
 - \frac{\partial \Phi }{\partial r} (r,t)  \approx r \frac{k_B T_e (t) }{m_i \sigma(t)} 
$$ 

The dimensionless equations are very similar to the self similar ones used in clusters physics \citep{louis1991}. The selfsimilar ionic evolution is easy to determined and verify:
\citep{robicheaux2002,robicheaux2003}:
 \begin{equation}
u_i (r,t) \approx   t r \frac{{v_0^2}{\sigma_0^2}}{ 1 + \frac{v_0^2}{\sigma_0^2} t^2  } 
\label{u_self_sim}
\end{equation}
for the ion mass transport velocity, which is merely equation (\ref{expansion}).  \citealt{dorozhkina1998} have demonstrated
 that this result is in fact more general.
 Using Vlasov equations for $f_e$ and $f_i$, assuming a quadratic form for the potential, a self similar homological evolution, quasineutrality and equation (\ref{equ_self_sim}) they also found equation (\ref{u_self_sim}).  
 In fact  this result has been recently improved and the most general form for a self similar quasineutral collisionless plasma expansion into vacuum, without any other assumption is:
 \citep{kovalev2003}:
\begin{equation}
f_f = f_f (I_f)\ ; \ I_f = \frac{ \vec v^2 + \Omega^2 (\vec r - \vec v t)^2}{2} + \frac{q_f}{m_f} 
\phi \left( \frac{\vec r}{\sqrt{1+\Omega^2 t^2} } \right) 
\label{Schwrd_bolt}
\end{equation}
 Using $\Omega = \frac{v_0^2}{\sigma_0^2}$ we recover
 equations (\ref{u_self_sim}) and (\ref{expansion}) which are then shown to be the most general self similar solutions for a quasineutral plasma.

 \subsection{Temperature evolution}
 
One important question concerns the temperature evolution in this self similar solution.
The energy conservation comes from the time derivative of the square of equation (\ref{Eq_Poisson}) (Ampere's law):
$$
\frac{\partial w^{\rm tot}}{\partial t} = \frac{\partial \Phi}{\partial t} ( \rho_e u_e + \tilde \rho_i u_i )
$$
 added to equation (\ref{ene_cons}) and spatially integrated:
 \begin{eqnarray}
\frac{3}{2} k_B \bar T_e (0) &= &\frac{3}{2} k_B \bar T_e (t) + \frac{1}{2} m_i \int_0^\infty u_i^2(r,t) n_i(r,t) 4 \pi r^2 d r \nonumber
\\
&= &\frac{3}{2} k_B \bar T_e (t)+ \frac{3}{2} m_i v_0^2  \left( \frac{t v_0 }{ \sigma (t)} \right)^2
\nonumber
\end{eqnarray}
 where we did not take into account the negligible ionic kinetic energy $ T_i \ll  T_e^\gamma$ neither the ionic correlation energy \citep{pohl2004b,pohl2004c}
or the heating due to the three body recombinaison (see \citet{robicheaux2003} and equation (\ref{heat_rate})).
We see that during the plasma expansion an electronic adiabatic cooling occurs and the ionic distribution function 
becomes a Schwarzchild-Boltzmann one given by equation (\ref{Schwrd_bolt}).

\subsection{Analytic ion spike evolution}

The collisionless theory, we have just developed,  was based on several assumptions and 
has the usual drawback of all self similar solutions predicting unphysical results as, for instance, a velocity increasing without limit for $r$ going to infinity. Furthermore the theory
is only valid before the ionic spike appears.

We know from Monte Carlo \citep{robicheaux2003} and molecular  \citep{pohl2004c} simulations that the quasineutrality is violated and that the ionic front shows a density spike after few microseconds. 
Some theoretical prediction \citep{pohl2004c} predicts that the ionic spike, where the quasineutrality is violated, disappear. This result has to be related with the work on non-ultracold plasma where it can be shown that the ion front moved at velocity varying logarithmically in time \citep{mora2003}. 

 Following \citealt{kaplan2003}, who studied a non neutral plasma, we will describe the $r$-radius (with $N_i(r)$ ions) ion shell evolution. We assume that along a $r$-shell trajectory $r(r_i,t)$ starting at the initial point $r_i$, the total number of ions remains unchanged: $N_i(r(r_i,t))= N_i(r_i)$. 
The ion density is then given by $n_i(r,t) = n_i(r_i,t=0) \frac{\partial r_i}{\partial r} \frac{r_i^2}{r^2} $.
The condition that no particle trajectories cross each other is violated when the shock formation appears creating an infinite local ionic density. A natural expansion of the quasineutrality is $N_e(r)\approx N_i(r)$, we will therefore assume
$ N_e(r(r_i,t))= N_e(r_i)$. 
The ion Newton equation is:
$$
 m_i \frac{\partial^2 r}{\partial t^2} = \frac{q_i^2}{4 \pi \varepsilon_0} \frac{ N_i(r_i)-N_e(r_i) }{r^2}
$$
integrating $\frac{d r}{d t}$ times this equation leads to the conservation of energy equation  and to the analytical implicit solution:
\begin{eqnarray}
\lefteqn{ \frac{t}{r_i^{3/2}} \sqrt{2 \frac{q_i^2}{4 \pi \varepsilon_0} \frac{N_i(r_i)-N_e(r_i)}{m_i}}
= } \nonumber \\
& & \sqrt{  \frac{r}{r_i} \left( \frac{r}{r_i} -1 \right) } + \ln \left(  \sqrt{\frac{r}{r_i}} + \sqrt{\frac{r}{r_i} -1} \right)
\label{eq_spike}
\end{eqnarray} 
A typical result with gaussian $\sigma \approx 250\mu$m approximation for 
$ n_i$ and $n_e$ and with $N_i-N_e = 20000$
is given in figure \ref{fig:spike}.

\begin{figure}
\resizebox{0.45\textwidth}{0.28\textwidth}{
\includegraphics*[0cm,8.8cm][12.4cm,16.8cm]{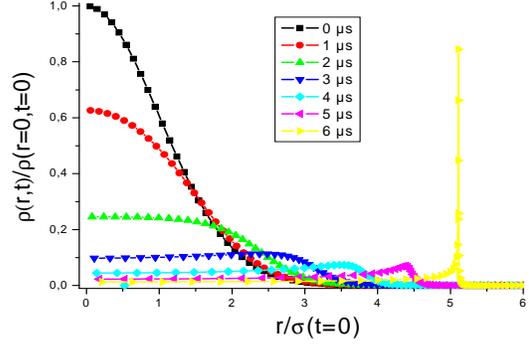}  
} 
\begin{center}
\caption{Ion density evolution based on equation (\ref{eq_spike}) for $\sigma \approx 250\mu$m  and $N_i-N_e = 20000$.}
\label{fig:spike}
\end{center}
\end{figure}

The typical ''Coulomb Explosion'' time scale is then
$t_{CE}= \sqrt{ \frac{ 4 \pi \varepsilon_0 }{q_e^2}} \sqrt{ \frac{m_i}{n_i^0-n_e^0} } 
\approx 3.5\,\mu$s where $n_i^0=n(r_i=0,t=0)$.
At early time $t\ll t_{CE}$ the evolution equation (\ref{eq_spike}) leads to an evolution
given by
$$
\frac{r}{r_i} = 1 + t^2 \frac{v_0^2}{r_i^2} \frac{3}{4} \sqrt{\frac{\pi}{2}} \frac{\sigma}{r_i} \frac{N_i(r_i)-N_e(r_i)}{N_*}
$$
where we have used equation (\ref{N_star}). 
With equation (\ref{ratio}), and with the gaussian approximation (\ref{Nb_ion})  this might be written (for $r=\sigma$): 
\begin{equation}
\frac{\sigma}{\sigma_0} \approx  1 + 0.28 \left(\frac{N_i}{N_*}\right)^{0.5} t^2 \frac{v_0^2}{\sigma_0^2}  
\label{expl_coul}
\end{equation}
 which is, with $\left(\frac{N_i}{N_*}\right)^{0.5}  $ on the order of $10$, is very similar to equation (\ref{expansion}) for early times. 
Even if the hypothesis $ N_e(r(r_i,t))= N_e(r_i)$ is probably worse than the one for ions. The final results should not be too badly affected by this assumption. Indeed, the only requirement for equation (\ref{expl_coul}) is the time independence of $N_i(\sigma(t)) - N_e(\sigma(t)) $ which is probably a better assumption.

\subsection{Closure relations for the electrons}

Let us go back to the electron evolution to find the  time dependent evolution equations for the electron distribution.
According to the ambipolar diffusion and due to the quasineutral plasma behavior we have (especially in the core region) $u \approx u_i$ so (in absolute value)
$$
 \frac{D u}{D t} \approx  \frac{D u_i}{D t} \approx \frac{m_e}{m_i} \frac{ \partial \Phi}{ \partial r} \ll \frac{1}{\rho} \frac{ \partial p}{ \partial r}
$$
Therefore we need to go at least one step further to close the moment equations for the electrons.
  For our isotropic  quasistationary evolution $3 p = \langle 2 \rangle$ and
 we  define 
(see \citep{louis1991})
 the velocity energy transport $v$ by 
 $5 p v = \langle 3 \rangle = \int  v_r v^2 f_m(\vec v) d^3 \vec v$ which is linked to the heat flux  (different from the energy flux across a constant radius sphere) $F= 5 p(v-u)/3 = L/(4\pi r^2)$ where $L$ is the luminosity.
 The hydrodynamical equation becomes with  $M_i^* = m_e N_i/M_0$:
 \begin{equation}
  \frac{\partial \ln p^*}{\partial \ln r^*} - \lambda	\frac{\rho^*}{p^*} \frac{M^* - M_i^*}{r^*} =0 
	 \end{equation}
	 where $\lambda=N_e/(N_i-N_e)$ and we have assumed $p_0 = (-G') (N_i-N_e) m_e \rho_0/r_0 = \rho_0 {{\sigma_v}_0}^2$. $\lambda=-1$ is the case without any ions and, in this case, equations are identical to the cluster case.
	 
The energy transport equation becomes
 \begin{eqnarray}
	0 &=& \frac{\partial p}{\partial t} + \frac{5}{3} {\rm div} (p v) +  \frac{2}{3} u \rho \frac{\partial \Phi}{\partial r} \\
	0 & = & \frac{\partial \ln p^*}{\partial t^*} +	\frac{5 v^*}{3 r^*} \frac{\partial \ln p^*}{\partial \ln r^*} +
	 \frac{5}{3} \frac{\partial v^*/r^*}{\partial \ln r^*} + 5 \frac{v^*}{r^*} - \frac{2}{3} \lambda \frac{u^* \rho^*}{p^*} \frac{M^*-M_i^*}{{r^*}^2} \nonumber
\end{eqnarray}
which can be written in a  first thermodynamical law form:
 \begin{equation}
		\frac{\partial L}{\partial r}  =  - 4 \pi r^2 \rho \left\{ 
		\frac{ D }{D t} \frac{3 k_B T_e}{2 m} + p 	\frac{ D }{D t} \frac{1 }{\rho}
		\right \} \label{thermo}
\end{equation}

 The most delicate assumption is the form of the closure relation.
A simple one is the thermal conductivity assumption 
 given by \citep{lynden1980,louis1991}:
\item The thermal conductivity closure relation

 \begin{equation}
	F  =  - K	\frac{ \partial T }{\partial r} = -  \Lambda_c	\frac{ \partial  \sigma_v^2 }{\partial r}     \label{heat_flux}
\end{equation}
where $K= k_B \Lambda_c/m_e $ is the thermal conductivity coefficient. The theory of heat flux in gases indicates that $\Lambda_c$ should be on the order of $\rho l^2/t_{\rm col}$. Where $l$ is the mean free path and $t_{\rm col}$ is  the time between collisions better to be 
 taken as the relaxation time $t_e$ (with velocity $\sigma_v$) see  \citep{spitzer1987,lynden1980} for the cluster case and \citep{majumdar1971} for the plasma case. 
 When $\Lambda$ is not too small,
the mean free path $l \approx 3 t_e \sigma_v$ (typically $2$ mm) is comparable or larger than the size of the sample $ \sigma $. 
This indicates that
the electrons make severals orbits before colliding with neighbors.  Hence, it is better to use for $l$ a typical radial distance $ l_r $ between encounters given for instance when the mean velocity $\sigma_v$ is the circular velocity $ l_r \sqrt{4 \pi (-G') \rho/3}  $ defined for local homogeneous system. 
Using dimensionless values we can see that $
 t_0 ={t_e}_0 = 9 {\sigma_v}_0^3 / (16 \sqrt{\pi} G'^2 m_e \rho^0 \ln \Lambda) $ and
 $l_0^2 = {l_r}_0^2 = \frac{3 {{\sigma_v}_0}^2 }{4 \pi (-G') \rho_0} =  \frac{3}{\lambda} r_0^2$.
This leads to formula:
\begin{equation}
\Lambda_c \propto  \frac{\rho l_r^2}{t_e} =   \frac{4}{9 \sqrt{\pi}} 3 (-G') m_e  \frac{\rho }{\sigma_v} \ln \Lambda
\label{heat_flux2}
\end{equation}
where the dimensionless proportionality factor should be on the order unity
(nearly $0.4 =   \frac{4}{9 \sqrt{\pi}} C$ in \citet{louis1990}).

 \end{itemize}

 The phenomenological heat flux formula (\ref{heat_flux}) is based on the assumption $l>\sigma$ which is not always verified, especially for small Coulomb logarithm value. 
 Furthermore the fact we use only local variable for the thermal conductivity equation (with $v_0= r_0/t_0$):
 \begin{equation}
\frac{v^*}{r^*} - \frac{u^*}{r^*} + \frac{9}{5 \lambda} \frac{{p^*}^{-1/2} {\rho^*}^{1/2}}{ {r^*}^2} 
\frac{\partial (\ln p^* - \ln \rho^*)}{\partial \ln r^*} = 0
  \end{equation}
  might not be the best solution (see \citep{louis1991}). 
It is also possible to use the higher momentum equations, calculated  by \citealt{larson1970} and by \citealt{louis1990}, namely:
	\begin{equation}
 \frac{\partial \kappa }{\partial r} + p \frac{\partial \Phi }{\partial r} =0
\end{equation}
where $15 \kappa = \langle 4 \rangle  = \int_0^{+ \infty} 4 \pi v^6 f_m(v) d v$ and
	\begin{equation}
	\frac{\partial \kappa}{\partial t} + \frac{7}{3} {\rm div} (\kappa (2v-u)) +  \frac{2}{3} p \frac{\partial \Phi}{\partial r} =  - \frac{3}{5 t_e} \left( \kappa - p^2/\rho \right)
\end{equation} 

  We have shown that the equations are almost identical in cluster and plasma systems with an extra $\lambda$ factor in the plasma case.
We shall not fully solve the equation here but the classical \citet{henyey1964}-Newton-Raphson implicit difference relaxation method is probably well adapted.
In fact we shall not resolve here the gaseous equation because the original Fokker-Planck equation should give better results. 
Let us note that using 
Abel's transform of
$
\rho(\Phi(r)) = 2^{5/2}  \pi \int_{\Phi(r)}^\infty f_m(E) \sqrt{E-\Phi(r)} d E
$
it is possible to recover the phase-space density distribution function:
$$f_m(E) = \frac{1}{ \sqrt{8} \pi^2} \frac{\partial}{\partial E} \int_E^{+\infty} 
\frac{\partial \rho / \partial \Phi}{\sqrt{\Phi-E}} d \Phi
$$
This is the well known Eddington  formula (1916) \citep{binney1987}.

The simplest solution we might think is a self similar one for electrons as well as for ions which in fact lead to a  stationary solution.
Gravitating systems never form static homogeneous equilibrium. However,
plasmas contain both positive and negative charges, so they can form static equilibrium.
The  time scale for ion expansion is order of magnitude the electrons relaxation time toward equilibrium. So electron will reach equilibrium on a nanosecond time scale where the ionic motion could be considered as frozen. 
We might then first look for static equilibrium distributions.
Unfortunately equations (\ref{thermo})
 leads to $L$ constant and equations  (\ref{heat_flux}) and (\ref{heat_flux2}) to $L(r=0)=0$ and $\sigma_v$ constant. A stationary solution has no heat flux and is isothermal with a density distribution given by:
 $$
 \frac{d}{d r} \left( r^2 \frac{d \ln \rho}{d r}  \right) = -\frac{G' 4 \pi }{\sigma_v^2} r^2 (\rho -\tilde \rho_i)
 $$ 
 This leads to a density distribution with a $r^{-2}$ power law at large distance meaning an infinite mass.
 In fact such distribution present an attractive potential for electrons up to a point $r_t$ where the ion and electron density becomes equals and electrons evaporates leading to a repulsive potential for $r>r_t$.
 We already know from Monte Carlo simulations \citep{robicheaux2003} that
this isothermal solution is  correct for the
 core system but the  finite mass ''halo'' can only tend to be isothermal.

 Before going back to the Fokker-Planck equation, we can mention that it is possible to take into account the three-body collisions through the heating term (\ref{heat_rate}) in the kinetic energy transport equation. We should also use equations (\ref{eqn_rate_MC}) and (\ref{Ryd_en_dist}) to give a full picture of free and bound electron distribution.


\section{Orbit average of the Fokker-Planck equation}
\label{sol_FP}

\subsection{Truncated assumption}

If the system is isolated the maximum energy 
above which one the particles are extracted from the sample
is $E_t=\Phi(r=\infty)$. But, the system can be truncated at a given (apocenter) distance $r_t$ and  the truncated energy 
 is then
$E_t=\Phi(r=r_t)$. 
In a isotropic Fokker-Planck equation one has no other choice but to use the energy as a criterion for escape. For cluster physics the  truncation  in energy is a good approximation even if it is known not to be  a sufficient condition to describe accurately the physics \citep{kim1999}. For cluster the differential forces, produced by the galaxy, result in two saddle points (distance from the cluster center which is then proportional to the cube of the mass) through which stars can pass over into the galactical field.
For plasma physics under magnetic field a velocity truncation is probably another good choice due to the velocity dependent force. But, the experiment can use an homogeneous electric field (see figure \ref{fig:trap}) $F \vec e_y$ along the $y$ axis. In this case the truncation arise only from a single saddle point at $y_0=r_t$ abscissa along the $y$ axis where
\begin{equation}
F = - \frac{\partial \phi}{\partial r} (y_0) = - \frac{m_e}{q_e} \frac{\partial \Phi}{\partial r} (y_0)  \label{radial_t}
\end{equation}
Therefore the truncation is not at all spherically symmetric and the hypothesis is probably worse 
 for plasma than for tidally limited cluster. 
  Nevertheless, under ergodic assumption, the hypothesis of truncation in energy might be a
  good one and will be used hereafter.
There is always a stray electric or magnetic field in the experiment. The system is therefore never perfectly isolated and cannot extend to infinity without touching some electric grids (see figure \ref{fig:trap}). Furthermore, an external electric field $F$ is sometimes constantly applied (as done by \citealt{kulin2000}) to be able to accelerate electrons (or ions) toward a detector and to be able to monitor the behavior of the sample. In such experimental conditions equations (\ref{radial_t}), (\ref{sol_Poisson}) and (\ref{Nb_ion}) imply
\begin{eqnarray}
F &\approx&
 G' \frac{m_e^2}{q_e}  \frac{N_e-N_i 
 \left(1- \sqrt{\frac{2 r_t^2}{\pi \sigma^2} } e^{-\frac{r_t^2}{2 \sigma^2}} \right) }{r_t^2} \ \nonumber
  \\
  & \approx& G'm_e  \frac{N_e-N_i}{r_t^2} =
\frac{q_e^2}{4 \pi \varepsilon_0}  \frac{N_e-N_i}{r_t^2}
   \label{eq_radius}
  \end{eqnarray}
  
This experimental electric field can be experimentally  used to extract electron trapped in an ionic cloud in order to simulate the tidal escape from globular cluster under external galaxy gravitational attraction. 
Typical value, before the expansion, are $r_t\approx 1.5\,$mm$ \approx 6 \sigma$ for $F=1\ $V/m.
We can see here that $r_t$ is proportional to the square of the mass and not to the cube of the mass as in the star cluster case under tidal effect of the host galaxy \citep{spitzer1987}.
 Assuming the system reacts on the external field we will neglect its role inside the system simply summarize it by the $E_t$ value. This assumption is better for a small external field.
 In our experiment the
magnetic field gradient ($B' = 1.5\,$mT/cm) is not turned off and it ejects electrons  for $r>
 r_t=12\sigma$ (value estimated for $T_e \approx 100\ $K). This value is similar (see equation (\ref{eq_radius}) to the radius found for electric field on the order of $F=2  \,$mV/cm which is indeed a typical value for uncontrolled straight electric field.
     As already discussed, 
the mean free path is usually large compared to the sample size, or similarly the orbital timescale at $r_t$ is negligible compared with the relaxation time. Thus, we will assume that the electrons with energy $E>E_t$ would be lost  almost instantaneously. We know this criterion is strongly violated in the star cluster case \citep{takahashi1998} but this is nevertheless a reasonable choice,
yielding to $f(E)=0$ for $E>E_t$ and by continuity 
\begin{equation}
f(E_t)=0 \label{continuity}
\end{equation}

The zero probability presence at the border is opposite to the conclusion obtain in ultra-cold atomic system interacting through van der Walls atomic interaction potential. Here the Coulomb (or Newton) interaction acts at very long distance and encounters are ''gentle'' leading to a small continuous variation in energy during collisions. On the contrary 
the van der Walls interaction is at so short range that only the strong collisions with a single encounters are involved. 
Consequently,
in the route toward quantum degenerate dilute ultra-cold atomic Bose Einstein Condensates the  evaporation (it is more an ejection process but this ''atomic'' terminology is well known) process leads to truncated maxwellian distribution with non zero probability presence at the truncated border.
Despites the differences similar strategies are used in stars system and in atomic system :  superstar strategies  in 
   cluster code  and macro-atoms strategies used in evaporative cooling
  models of ultra-cold neutral atomic cloud in route to Bose-Einstein condensation \citep{berg-sorensen1997,pinke1998,tol2004}.

\subsection{Orbit averaging}

 Using $v = \sqrt{2(E-\Phi)}$ we define
 the volume $\tau$ of phase space with energy less than $E$ and 
the isotropic average $q$ of the radial action by:
\begin{equation}
q(E,t) = \frac{1}{3} \int_{0}^{\Phi^{-1}(E)}  (2(E-\Phi(r,t))^{3/2} r^2 d r = \frac{1}{16 \pi^2} \tau(E,t)
\label{orbit_aver}
\end{equation}
For our quasi gaussian system we found, from equation (\ref{phi_gauss}) that a good approximation (within 10\% accuracy) for $q$ is:
\begin{equation}
q(E) \propto (E_t-E)(E-E_0)^3
\label{phas_gauss}
\end{equation}
This is not surprising because we are between
 a square potential, where $\Phi=0$ and $q\propto E^{3/2}$, and a quadratic potential, where
$\Phi = \Phi(0) + \frac{1}{2} \omega^2 r^2 $ and
$
\tau =\frac{4 \pi^3}{3} (E-E_0)^3
$.
We will also sometimes use $\frac{\partial q}{\partial E}=\int_{0}^{\Phi^{-1}(E)} r^2 \sqrt{2(E-\Phi(r,t))} d r$ which is
 the isotropic average of the period and is $1/(16 \pi^2)$ times the classical microcanonical phase space density function.
Indeed, on the contrary to the ultra-cold atomic system case, the plasma is far from quantum degeneracy - the Fermi temperature is on the order of tens of microKelvin which is orders of magnitude below the plasma temperature - so we can use the classical density phase distribution function which is the same than the quantum one but multiplied by the fundamental phase volume $h^6$ ($h$ is the Planck's constant). 

Multiplying the Fokker-Planck equation (\ref{FP_tot}) by the delta function $\delta(E-(v^2/2+\Phi))$ 
and integrating over the phase space leads to the
orbital average Fokker-Planck equation \citep{henon1961,spitzer1987}
\begin{eqnarray} 
	\lefteqn{ \frac{\partial q}{\partial E} \frac{\partial f}{\partial t} -  \frac{\partial q}{\partial t} \frac{\partial f}{\partial E}  =  
	4 \pi \Gamma  \frac{\partial }{\partial E} \Big[
	f \int_{E_0}^E f' \frac{\partial q'}{\partial E'} d E' + } \nonumber \\
	& & \frac{\partial f}{\partial E}
	\left\{
	\int_{E_0}^E f'  q' d E' + q \int_E^{E_t} f' d E'
	\right\}
	\Big] \label{FP_E} \\
	\left(  \frac{\partial f}{\partial t} \right)_\tau & = & 	\left(  \frac{\partial \Pi}{\partial \tau} \right)_t
	\label{FP_E_Nice}
\end{eqnarray}
where $E_0=\Phi(0,t)$. Equation (\ref{FP_E_Nice}) is merely equation (\ref{FP_E}) divided by $ \partial \tau/\partial E$ \citep{inagaki1990}. The flux through phase space $\Pi$ can be written in another nice form:
$$
\Pi = 4 \pi \Gamma  \int_{E_0}^{E_t} f f' {\rm Min}(\tau,\tau') \left(
\frac{\partial \ln (f)}{\partial E} - \frac{\partial \ln (f')}{\partial E'} 
\right)
d E'
$$
which define
 the generalized concept of temperature $T_e^G(E)$  by
$$-\frac{1}{m_e} \frac{\partial \ln f}{\partial E} = \frac{1}{k_B T_e^G (E) }$$
The formulas can be checked with the use of
 a square potential ($\Phi=0$ for $r<r_t$).   We have, in this case as in free space, $q\propto E^{3/2}$, $E = v^2/2$ and equation (\ref{FP_E}) recover exactly the Fokker-Planck equation (\ref{FP_tot}).

Equation (\ref{FP_E}) is called the orbit average Fokker-Planck equation and holds when
the relaxation time is longer than the crossing time or the orbital time. As
previously mentioned we can  experimentally violate this condition  (for very cold electrons or very large and dense sample) but usually this assumption holds. 
Thus the distribution function will evolve  slowly compared to the orbital period of each electron.
In this slowly time variable potential (orbit-averaged assumption) 
$$dE/dt \approx \int dr r^2 v (\partial \Phi / \partial t)/\int dr r^2 v =
\frac{\partial q / \partial t}{\partial q / \partial E}
$$ is the mean value of  $\frac{\partial \Phi}{\partial t}$, hence 
$q$ and $\tau$  are adiabatic invariant: $\frac{d q}{d t} = \frac{d \tau}{d t} = 0$ 
 and therefore $f(E,t)= f(\tau)$ \citep{binney1987,inagaki1990}.

 Methods of calculation, developed for stars clusters, can then be very easily adapted. For instance the Chang and Cooper (1970) finite-differencing scheme used in Cohn's method \citep{cohn1979} is perfectly adapted. Indeed, the first step of the method is to advance $f$ in time $\Phi$ being held fixed. The Fokker-Planck equations are identical for both systems so no change has to be done in this code step
 going from stars system to ultra-cold plasma system. The second, easier step, is to
advance  $\Phi$  by solving the Poisson's equation with $f$ being fixed as a function of the phase volume adiabatic invariant $\tau$. 
Other two step methods can also be used, as the one developed by  \citealt{takahashi1993} from variational principle. 
We hope that the simplicity of this project would lead to rapid use of Fokker-Planck code in ultracold plasmas physics.

\subsection{Evaporation and ejection}
Electrons can escape from the plasma by {\em ejection} (through strong collisions with a single encounters) or be  {\em  evaporation} (through series of weaker distant encounters) \citep{binney1987}. 

The evaporation rate has a long history in stars dynamics (e.g. \citep{johnstone1993}).
Let us first formulate, for our plasma system, the
\citet{henon1960} 
 paradox known in clusters physics:  in the case of isolated system ($r_t=\infty$) an electron with energy $E\approx E_t$ spend most of its time far from the center and suffer very few encounters. 
 Its change energy rate goes to zero when $E$ approach $E_t$ so the electron never escape in this diffusion type (Fokker-Planck) picture. Thus, the orbit average Fokker-Planck equation, with the same relaxation time for all electrons, do not hold for outermost ''halo'' electrons \citep{spitzer1972}. 
The fact that halo electrons play a role has been in fact experimentally sketched using Radio Frequency heating of electrons. Indeed, \citealt{li2004} used RF electric field to ''shake'' the trapped electrons and speed up the ionization of Rydberg atoms inside the plasma. They have noticed that the process is less efficient for large RF amplitude  due to the fact the forced RF electron oscillations drive electrons outside the cloud from the center reducing the amount of time the electron could collide with Rydberg atoms. This experiment indicates that when electrons are in the halo they do not collide that much.

From H\'enon's paradox we can say that
the electron can escape
from an isolated system only from {\em ejection} due to strong single close encounters. The rate has been calculated by \citealt{henon1960} in the cluster case. By analogy using equation (\ref{replacement}) we could then assume:
\begin{eqnarray}
\lefteqn{
\frac{d N_e}{d t} = \frac{2}{3} (16 G' m_e^2 \pi^2)^2 \int_0^\infty r^2 d r \times } \nonumber \\
& & \int\!\!\! \int  d E d E'
\frac{(E+E'-\Phi-E_t)^{3/2}}{(E_t-E)^2} f(E) f(E')  
\end{eqnarray}
where the integration hold for $E'\geq E_0$, $E_t \geq E\geq E_0$ and $E+E'\geq \Phi+E_t$.
Performing the  integration over $E'$ leads to the loss 
rate of electrons with energy $E$.

The case of a truncated system ($r_t<\infty$) is different and the evaporation process is a diffusion in velocity space process well described by the Fokker-Planck equation.
The number $N(E)$ of electrons with energy less than $E$ is 
\begin{equation}
N(E,t) = \int_{E_0(t)}^E f(E',t) \frac{\partial \tau}{\partial E}(E',t) d E'
\label{equ_N}
\end{equation}
 time derivative straightforwardly leads to 
$$
\frac{ \partial N}{\partial t} ( E) = f(\tilde E) \frac{d   E}{d t}  \frac{ \partial \tau}{\partial E}( E) +
\int_{E_0}^{  E} \frac{\partial \Pi}{\partial E} (E') d E' 
+   f( E) \frac{ \partial \tau}{\partial t}( E)  
$$

When $E=E_t$ (or $\infty$), $N=N_e$ and the  first (and third) ''spilling'' terms  are null (see equation (\ref{continuity})). The second one can easily be calculated
using the general form 
$$
f(E,t) =f (E,t) Y(E_t(t)-E)
$$
where Y is the Heaviside function. 
With $\frac{dY}{dE} (E_t-E)=-\delta (E_t-E)$ we found that
\begin{equation}
\frac{ d N_e}{d t} = - 12 \pi \Gamma \frac{\partial f}{\partial E}(E_t) \int_{E_0}^{E_t} f \tau
\label{loss_rate} 
\end{equation}
which is similar to formula given by \citep{wiyanto1985} but for an unknown reason with a different numerical factor ( $3  \Gamma/(32 \pi) $ in their case).

To conclude we could note that
the evaporation process can also be taken into account by the gaseous model. Following
\citealt{heggie1998} we can for instance use an additional equation based on the velocity escape at radius $r$ law:
$$
\frac{\partial \rho}{\partial t} \propto \frac{r}{\sqrt{E_t-\Phi(r)}} \rho
$$
The flux conservation requires that the density distribution of the escape electrons should follow $\rho_{\rm esc} \propto r^{-2}$ \citep{spitzer1972}.



For a non isolated case
the particles acquire energy marginally in excess of escape energy $E_t=\Phi(\infty)-G' M m_e/r_t$.  The  energy loss rate is therefore $\rmn{d} H=  -G'  m_e \rmn{d} M /r_t $ and can be seen as the results of the work done on the plasma by the external electric field. 
 

\subsection{Stationary solutions}
 
   There is many ways to derive the quasi-stationary solution of the Fokker-Planck equation where the left hand side of the orbit average Fokker-Planck equation verifies $d f/d t \approx 0$ so  $\Pi= \Pi_0$ is constant. Following \citealt{king1965} we write a first order linear differential equation form
\begin{eqnarray}
\frac{\Pi_0}{4\pi \Gamma}  & = & f \int_{E_0}^{E} f'
 \frac{\partial  \tau}{\partial E'} 
 +   
 \frac{\partial  f}{\partial E} 
\left(
	\int_{E_0}^E f'  \tau' d E' + \tau \int_E^{E_t} f' d E'
	\right) \nonumber \\
& \equiv &  f N_{\tau}
 +   
 \frac{\partial  f}{\partial E} 
 H_{\tau} \label{FP_first_order}
\end{eqnarray}
 $\Pi_0$ is  proportional to the evaporation rate, and $N_\tau$ is proportional to $N$ through equation (\ref{equ_N}). 
If $\Pi_0 =0$ it is easy to find that the only solution is a maxwellian one  \citep{henon1960} which is unphysical because non-truncated.

\subsection{Kramers-Michie-King solutions}

With the same assumption used to derive equation (\ref{thermal_bath}) namely a thermal bath  Maxwellian 
 assumption form for $f'$, we could go one step further and simplify the  Fokker-Planck equation. In the high energy limit ($ m_e E = {\cal{E}} \gg k_B T_e$) it reduces to an equation, first derived by Kramer (1940) for the brownian motion \citep{melnikov1991}: 
\begin{equation}
\frac{{\rm d} f}{{\rm d} t} \approx \frac{ \Gamma n_e(r)}{2 v} \frac{\partial}{\partial {\cal E} }\left[ \frac{f}{k_B T_e} + \frac{\partial f}{\partial {\cal E} }\right]. \label{Kramers}
\end{equation}
Our quasi-equilibrium case, $\frac{{\rm d} f}{{\rm d} t} \approx 0$, 
 immediately leads to the Kramer (1940)-Michie (1963)-King (1965) type quasi-equilibrium \citep{king1966} for the 
phase-space density function $f$ distribution :
$$f \propto (e^{-\frac{{\cal E}}{k_B T_e^K}}-e^{-\frac{{\cal E}_t}{k_B T_e^K}})$$  
We define what we shall call the Kramers-King's temperature $T_e^K$.

The electron density $n_e(r,t)=\int_0^{+ \infty} f(r,v,t) 4\pi v^2 {\rm d} v$  can then be analytically calculated from the King's distribution:
\begin{equation}
    n_e \propto e^{\eta_t}{\rm Erf}(\sqrt{\eta_t})-\sqrt{\frac{4 \eta_t }{\pi}}\left(1+\frac{2}{3}\eta_t \right) \equiv F^K(\eta_t)
    \label{density_ne_eq}
\end{equation}
where the proportionality factor is obviously $n_e^0/F^K(\eta)$ where $n_e^0$ is the electron density at the cloud center and
$\eta=\eta_t(r=0)$ (notation chosen to be identical to the one used in BEC evaporations theories). $\eta_t(r,t)=\frac{{\cal E}_t(t)- q_e \phi(r,t)}{k_B T_e^K(t)}$ is solution of the self-consistent Poisson
equation:
  \begin{equation}
    \frac{1}{r} \frac{\partial^2}{\partial r^2} (r \eta_t(r,t)) = \frac{q_e}{k_B T_e^K(t)} \left[
    \frac{q_i n_i(r,t) + q_e n_e (r,t)}{\varepsilon_0}
    \right] 
    \label{kramers_king}
\end{equation}
Similarly, with zero velocity of mass transport ($ k_B T_e= m_e \sigma_v^2 $), we have the important relation
\begin{equation}
 k_B T_e(r) = k_B T_e^K  \left[ 1 
 -   \frac{ \frac{8}{15 \sqrt{\pi} }\eta_t(r)^{5/2}  }{ F^K(\eta_t(r))} \right] \approx k_B T_e^K(t) {\rm Erf} (0.22 \eta_t)
    \label{TKing_vs_Te}
\end{equation}
with $N_e \bar T_e = \int_0^\infty 4 \pi r^2 n_e(r) T_e (r)$.
 We could noticed that, for equations (\ref{density_ne_eq}) and (\ref{TKing_vs_Te}), the adding factor to $e^x {\rm Erf} (\sqrt{x})$ is always the power series expansion for $e^x {\rm Erf} (\sqrt{x})$ about the point $0$. 

\begin{figure}
\resizebox{0.45\textwidth}{0.20\textwidth}{
\includegraphics*[0mm,95mm][236mm,181mm]{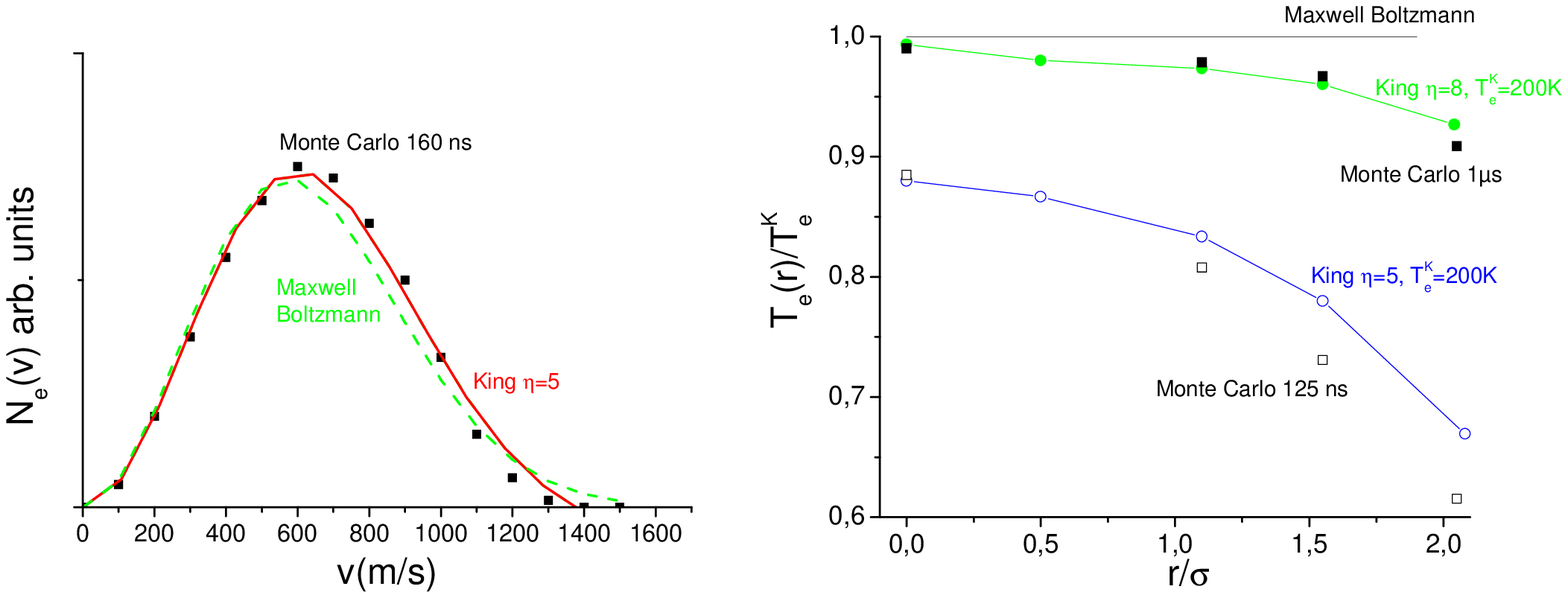} 
 } 
\begin{center}
\caption{Comparison between a Monte Carlo simulation, the King distribution and the maxwellian one. The left panel shows a comparison between the velocity distribution 
$N_e(v) \propto \int_0^{v} v'^2 f(v'^2/2+ \Phi(r_i) ) d v' $ taken at the $1/4$ shell such as $N_e(r_i)=N_e/4$.
The Monte Carlo calculation comes from \citealt{robicheaux2003} figure 1 after $160 \,$ns expansion time. The Kramer-King distribution is $\eta=5, T_e^K\approx 200\,$K.
The right panel shows the radial dependence of the temperature $T_e^r$ calculated by formula (\ref{TKing_vs_Te}) compare to \citealt{robicheaux2003} figure 2 for an expansion time of $125\,$ns and $500 \,$ns. The Maxwellian distribution is radially independent and is clearly not adapted.
}
\label{fig:King_vs_MC}
\end{center}
\end{figure}

At first approximation we can say that electrons are always in King's type quasi-equilibrium during the whole expansion of the ionic and electronic cloud.  Indeed, in figure \ref{fig:King_vs_MC},
using Figs. 1 and 2 Monte Carlo simulations  of  \citet{robicheaux2003}, we demonstrate that the King electron distribution is  better  than a pure Maxwellian one.

\begin{figure}
\resizebox{0.45\textwidth}{0.28\textwidth}{
\includegraphics*[0mm,97mm][141mm,191mm]{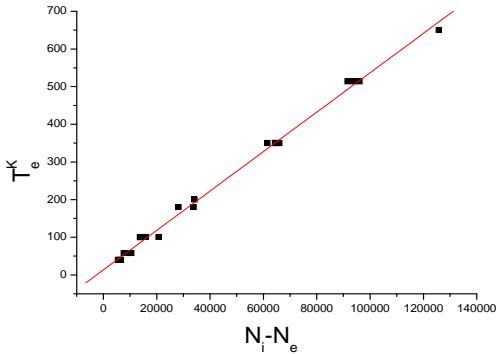}  
} 
\begin{center}
\caption{Comparison between numerical calculation and formula (\ref{temp_simpl}) the square data comes form several $N_e$, $N_i-N_e$ couples and different $\eta$ values ($3<\eta<15$) with $r_t=15-20$.}
\label{fig:simp_king}
\end{center}
\end{figure}

 The harmonic potential approximation in equation (\ref{kramers_king}) leads to:
\begin{equation}
k_B T_e^K  \approx \frac{ q_e^2}{3 \varepsilon_0} \sigma^2 (n_i^0-n_e^0)
\label{temp_king} 
\end{equation}
This equation is in fact well confirmed by
numerical resolution of equation (\ref{kramers_king}) for $10 \,$K$< T_e^K < 1000\, $K.
This expression has to be compared with equations (\ref{temp_gamma}) and (\ref{virial_res})  because the electron are not in a gaussian distribution so $n_e^0 \neq N_e/(2\pi \sigma^2)^{3/2}$. In fact, we found that
 $n_i(r)=n_e(r)$ for $r/\sigma\approx 3-4$ almost independently of all the other parameters and that
the electron  density distribution varies as $r^{-5/2}$ near infinity as expected.
 These results can be used to improved the naive formula extract form the Virial theorem. The numerical results
can be approximated, see figure \ref{fig:simp_king} by: 
\begin{equation}
1.9(\eta-2) k_B T_e^K \approx \sqrt{\frac{2}{\pi}} \frac{q_e^2}{ 4 \pi \varepsilon_0}  \frac{N_i-N_e}{\sigma}
\label{temp_simpl}
\end{equation}
This is close to the intuitive results:  the trapping depth $\eta k_B T_e^K$ has to be roughly equal to the trap depth,  calculated assuming a gaussian shape for ions and electrons
where $1.9(\eta-2)$ replace the naively $\eta$ expected value.

 \subsection{Second order solutions}

 A simple extension of our work would be to look at the evolution of the key parameter $\eta(t)$. 
An equation for the sole $\eta$ parameter can be derived  from the entropic variational principle \citep{takahashi1993}:
\begin{equation}
0= \int_{E_0}^{E_t} \left[
\left. \frac{\partial f }{\partial t} \right)_E \frac{\partial \ln f }{\partial \eta} \frac{\partial \tau }{\partial E}
+ \Pi \frac{\partial  }{\partial \eta} \left( \frac{\partial \ln f }{\partial E}  \right)
 \right] 
d E
\end{equation}
Another possible equation will be the one based on correct mass, energy and entropy evolution as done in cluster physics \citep{spitzer1987} or in ultra-cold atomics physics \citep{berg-sorensen1997,pinke1998}.

In the article we have only developed the isotropic case. Obviously another simple extension of our work will be to use anisotropic model.
 Similar Fokker-Planck equations holds for more general anisotropic system.   As a simple example we could indicate 
a natural  form for the electrons quasi-equilibrium distribution: the Michie's one used in anisotropic sample:
$$f\propto e^{J^2/J_0^2} (e^{-m_e E/k_B T_e^K}-1)$$ 
where  the second integral $J$ is the angular momentum. It is possible to average $f(E,J,t)$ over $J$ to restore the equation for $f(E) $\citep{cohn1979}.
In such Michie distribution the electron  density distribution varies as $(r_t-r)^{3/2} r^{-7/2}$ near $r_t$ \cite{binney1987}.

To study the general properties of the plasma it would be profitable to go one step further \citep{prata1971}. Indeed to study the velocity, or energy, distribution the first order is sufficient because these quantities are mainly determined by the ''maxwellian'' center where the collisions occurs. On the contrary, the escape rate is also determined by the velocity distributions in the whole system where $f$ is better approximated by the King distribution.

Using equation (\ref{continuity}) we solve the first order linear differential equation  (\ref{FP_first_order}):
\begin{equation}
f(E) =
 - \int_{E}^{E_t} \frac{\Pi_0}{4 \pi \Gamma H_\tau(E')} e^{\int_{E}^{E'} \frac{N_\tau}{H_\tau}} d E'
\label{second_order}
\end{equation} 
This gives the new $f$ function and then the news $\rho,\Phi,\tau$ functions from equations (\ref{sol_Poisson}) and (\ref{orbit_aver}). 
 Using  formulas (\ref{pot_gauss}) and (\ref{phas_gauss}) and assuming  the Kramers-King's  distribution 
leads to $10\%$ accuracy (for $1<\eta$ and for all possible $E$ values) approximation 
\begin{eqnarray}
 \int_{E_0}^{E} \frac{N_\tau}{H_\tau} &\propto & \frac{\eta}{1-e^{-\eta/2}} (E-E_0) \label{n_tau} \\
H_\tau & propto & f'_0  (E-E_0)^3 e^{(11 \frac{E_t-E}{E_t - E_0}+2\eta)/3} \eta^{1/3} \nonumber
\end{eqnarray}
The $(E-E_0)^3$ behavior near the core region $E\approx E_0$ leads to a diverging integral in equation (\ref{second_order}). 

The model looks pathological. But, with the King distribution, we can see that the energy flux is not zero at the center. By analogy with \citealt{henon1961}, we may say that the
energy flux
 is due to the Three-Body encounters.
The (two body) Fokker-Planck equation contain the fact the three body recombination does exist !
The energy flux emerging from the center is in fact realistic because it is supplied by binary formation. 
Therefore,
we need to put the three body collisions inside the Fokker-Planck equation.
 By analogy with the cluster case (e.d. \citet{takahashi1993}) 
 we must add an energy source term to the first-order diffusion coefficient \citep{spitzer1987} (2.79).
 Using the orbit averaging of the heating rate (\ref{heat_rate}) we need to replace the flux $\Pi$ in equation (\ref{FP_E_Nice}) by $	\tilde \Pi = \Pi - \tilde N f $ with :
\begin{equation}
  \tilde N = 16 \pi^2 \int_{0}^{\Phi^{-1}(E)}  \dot E (2(E-\Phi(r,t))^{1/2} r^2 d r 
  \label{TBR_FP}
\end{equation}
The only change is to change $\tilde N$ by $\tilde N+N_\tau$, so 
if $\tilde N$ is proportional to $N_\tau$ we recover, through equations (\ref{second_order}) and (\ref{n_tau}), the King solution. This is a strong indication of the good validity of this kind of solution as already proved by our comparison with Monte Carlo simulations.
In the cluster case the number of binary is always negligible and is usually not taken into account in this modified Fokker-Planck type equation.  Equation (\ref{heat_rate}) indicates that this assumption is not as good in a plasma sample, typically twenty ($100/5.4$) times worse. We then should add a loss term on the right side of equation (\ref{FP_E_Nice}). 
A better solution might be to use the orbit average master equation (see \citet{mansbach1969} (IV.1) or in \citet{goodman1993} (2.21)).
The rate constant is given by equation (\ref{eqn_rate_MC}) 
with $ ( k_B T_e)^2  K(E,E')= m_e^2 k(m_e (E-E_0) /(k_B T_e),m_e (E'-E_0) /(k_B T_e))$ :
\begin{equation}
\left. \frac{{\rm d} f}{{\rm d} t} \right)_{TBR} = n_e \int_{-\infty}^{+\infty} [K(E,E')f(E',t)-K(E',E)f(E,t)]
 {\rm d}E'
\end{equation}


We shall not develop here further on this discussion but we could mention that the Fokker-Planck equations with three body heating terms is similar to the self similar Fokker-Planck  equation without binary heating by \citealt{heggie1988} especially   with $\tilde N \propto q f$ and $\dot E $ constant.

\section{Evaporation experiments and electron temperature measurement}
\label{Temperature}

One of the most interesting part in the ultra-cold plasma system is its ability to be experimentally tested. We have already mentioned some experiments as the plasma expansion (through electron density measurement) or the TBR processes.  

Here, we would like to test the phase state distribution $f$. It could be  
experimentally tested for instance
 using  an electric pulse  to 
extract electrons \citep{vanhaecke2004,roberts2004} similarly to  ``runaway electron'' experiments \citep{kulsrud1973}. 
 A similar technique has been used in neutral atom Bose
Einstein Condensation (BEC) confined in a static external potential \citep{doyle1989}. The extraction should lead
to an instantaneous picture of the energy distribution of the electrons in the plasma. But  our case  is  complex
due to the fact the potential  depends on the
number of trapped particles  through  Poisson's equation. 

\subsection{Experimental determination of the  temperature.}

\begin{figure}
\resizebox{0.45\textwidth}{0.28\textwidth}{
\includegraphics*[4.5cm,4.5cm][22.5cm,17.0cm]{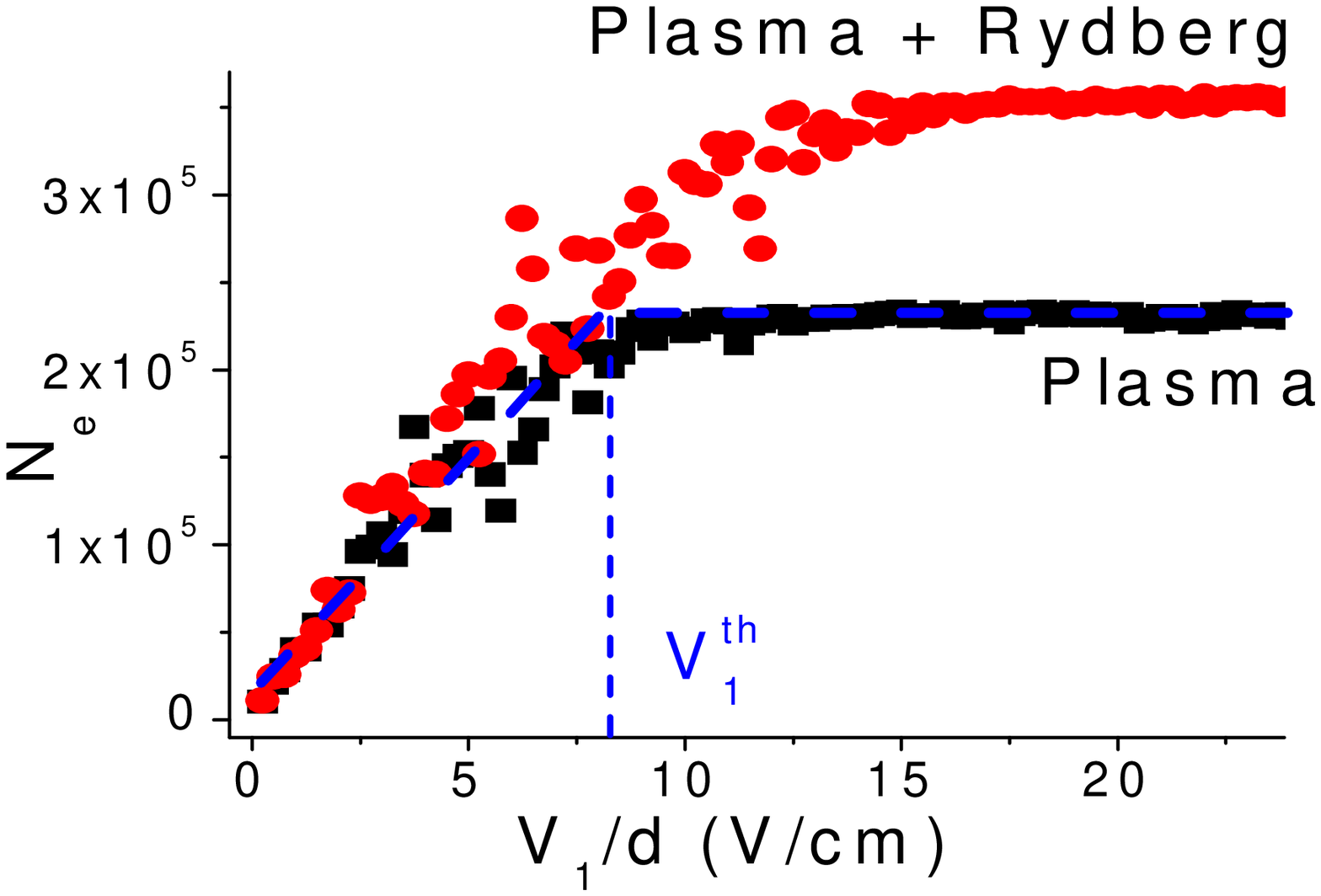} 
 } 
\begin{center}
\caption{Number of electrons ejected when varying the voltage $V$ applied  after $1\,\mu$s plasma expansion time. The dashed lines are guide for the eye to interpret equation (\ref{max_field_eq}). 
Each data have been calibrated to the average value of the total electron number.}
\label{fig:pulse}
\end{center}
\end{figure}

 Electronic temperature is a key parameter in plasma. If the thermal energy is lower than the
Coulomb interaction energy the plasma approaches the strongly coupled regime where correlation effects become
important \citep{pohl2004}. The basic idea we have used in \citep{vanhaecke2004} was to experimentally determine the temperature using a short voltage pulse $V$ to
extract electrons. The number of electrons ejected  by
the voltage $V$ is plotted in Fig. \ref{fig:pulse} both for the plasma only, and for the plasma plus Rydberg sample.

The first step of the theory is to use the threshold value $V_1^{\rm th}$ (see Fig. \ref{fig:pulse}) that is necessary to remove
all the free electrons to find $n_i^0$. Assuming there are always a few electrons with zero velocity in the 
Lagrange point where electrons are extracted from the
ion potential well,
the maximum electric field $F^{\rm th}$ created by the ionic space charge is:
\begin{equation}
 F^{\rm th} = \frac{V_1^{\rm th}}{d} \approx 2.38 \frac{q_i}{4 \pi \varepsilon_0} n_i^0 \sqrt{2 \sigma^2}
    \label{max_field_eq}
\end{equation}
Analysis of Fig. \ref{fig:pulse} and use of formula (\ref{max_field_eq}) lead to knowledge of $n_i^0$ at few
percent accuracy (assuming $\sigma$ is known exactly).
 Experimentally, one main uncertainty is the imprecise determination of $N_e$ due to laser
fluctuations and poor calibration of the charged particle detector (Micro Channels Plate MCP) detector. 
Our data were interpreted using a Kramer-Michie-King electron distribution.
The  last step of the theory is based on the solution of Poisson's equation with $n_i^0$ known
from equation (\ref{max_field_eq}). Because some electrons are removed even for small $V_1$ values, $n_e(r)$ this indicates that $r_t \approx \infty$. In order to take into account our magnetic and stray electric field we shall assume $r_t \approx 12 \sigma$. 
Furthermore, the total number of calculated electrons
$N_e(t)=\int_0^{+
\infty} n_e(r,t) 4\pi r^2 {\rm d} r$ should reproduce the observed number $ N_e$. This determines the values for the two remaining
parameters
$n_e^0$ and $T_e^K$ as a function of the unknown $\eta$ parameter through formula (\ref{temp_simpl}).
Unfortunately, 
we were too sensitive to the exact value of $\eta$ to be able to give an absolute value for the plasma temperature. Figure \ref{fig:King_vs_MC} and
Fig. 10 of  \citep{kuzmin2002b} show that $\eta\approx 8$ seems a reasonable choice to interpret the data given in figure \ref{fig:pulse}. Even with this
not fully satisfactory assumption we have used in \citealt{vanhaecke2004} the theory to give relative results. One conclusion was that the temperature should not increase or decrease by more than a factor $5$ when Rydberg atoms are added into a plasma. 

\begin{figure}
\resizebox{0.45\textwidth}{0.28\textwidth}{
\includegraphics*[0.6cm,0.6cm][10.2cm,9.0cm]{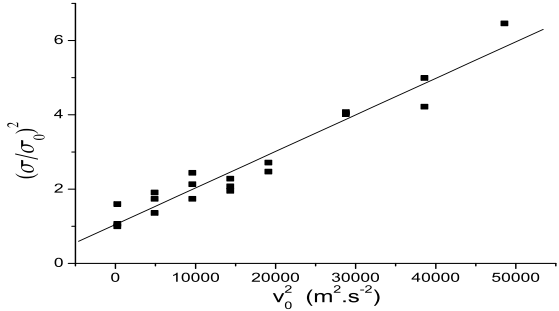} 
 } 
\begin{center}
\caption{Plasma size extract from expression  $\frac{\sigma}{\sigma_0}= \frac{N_i}{N_i^0} \frac{ V_1^{\rm th} } { V_0^{\rm th}} $ compared to $v_0^2 \approx k_B T_e^\gamma/m_i$ which is known from the laser ionization wavelength.
 $\sigma_0,N_i^0,V_0^{\rm th}$ are respectively the sample size (assumed to be frozen), the ion number and the threshold voltage for the laser sets at the ionization threshold ($T_e^\gamma \approx 0$). $\sigma,N_i,V_1^{\rm th}$ are respectively the sample size, the ion number and the threshold voltage for different laser wavelength taken after $2\ \mu$s of expansion time. The linear fit is $\sigma^2 \approx \sigma_0^2 + 10^{-4} v_0^2 $.}
\label{fig:thr_test}
\end{center}
\end{figure}

The theory presented here is based on several assumption which have to be tested. First,
to check the validity of formula (\ref{max_field_eq}) we have compared, after $2\ \mu$s of expansion time, the plasma size obtained by formula (\ref{max_field_eq}) to the one given by the expression (\ref{expansion}) for different plasma expansion velocity $v_0$. 
Using $n_i^0=\frac{N_i}{(2\pi \sigma^2)^{3/2}}$, equation (\ref{max_field_eq}) leads to  $\sigma \propto N_i V_1^{\rm th} $. $N_i$ is known from MCP The experimental results are depicted in figure \ref{fig:thr_test} where  $v_0^2 \approx k_B T_e^\gamma/m_i $ is varied by changing the laser ionization wavelength.
 Fitting results are in agreement with formula (\ref{expansion}) for $\sigma_0 \approx 200\ \mu$m which is very close to the expected MOT size. Therefore, we have here a strong indication concerning the validity of formula  (\ref{max_field_eq}).

\begin{figure}
\resizebox{0.45\textwidth}{0.28\textwidth}{
\includegraphics*[0.6cm,0.6cm][10.5cm,9.0cm]{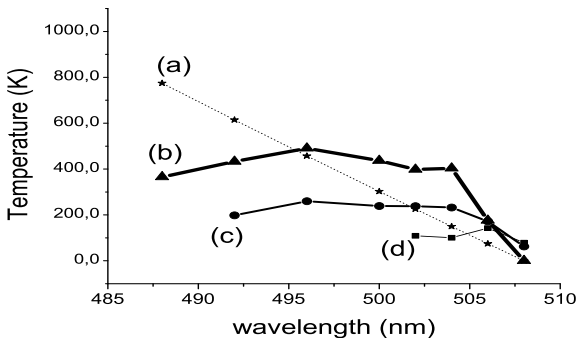} 
 } 
\begin{center}
\caption{Kramers-King temperature  $T_e^K$, extracts from
 equation (\ref{temp_simpl}) (with $\eta=7$), versus laser wavelength. 
(a) The photoionization electronic temperature $T_e^\gamma$ defined by:
$ 3 k_B  T_e^\gamma /2$  equals to the difference between the photon energy and the ionization threshold (near $508\ $nm). (b) data taken with full laser power (large $N_i$ value), (c) with  $5$ times lower power than in (b) (medium $N_i$ number, and (d) with  $15$ times lower power than in (b)
(small $N_i$ value).}
\label{fig:MK_test}
\end{center}
\end{figure}

Second,
we have compared formula (\ref{temp_simpl}), which gives $T_e^K$, with $T_e^\gamma$. 
$T_e^K$ has been experimentally extracted using  formula (\ref{temp_simpl}), with $\eta=7$, $\sigma$ is obtained using the results in figure \ref{fig:thr_test},  $N_i$ is given by  equation (\ref{temp_simpl} knowing  the voltage threshold value $V_1^{\rm th}$.
Results in figure (\ref{fig:MK_test}), for three different ion numbers, seems to indicate that $T_e^K$ is close to $T_e^\gamma$ and that $\eta=7$ is a reasonable value, after $2 \ \mu$s of expansion time, but has probably to be adapted for each ion number case.

Up to now we have just used one single point of the full curve given in figure \ref{fig:pulse}, a more complete study should yield to a determination of the $\eta$ value.
The full process is beyond the scope of the article but might be tackled with the ''violent relaxation'' theory \citep{ziegler1989,wiechen1994,chavanis1998}. Indeed,
 even in an impulse approximation with Fokker-Planck-Vlasov collisionless equations, or adiabatic approximation in gaseous equations, the violent ejection is a very complex system. 
At the present experimental status we have only investigated a global parameter namely
 the electronic temperature of the plasma.

\subsection{Other methods}

Many other methods may be uses to determined the plasma temperature. 
\begin{itemize}
\item \citealt{walz2004} have compared an experimental Rydberg binding energy distribution 
with calculated outcomes of inelastic collisions between
Rydberg atoms and electrons in the plasma taken from equations (\ref{eqn_rate_MC}).
They conclude that it is very likely that
 electron-Rydberg atom collisions cause most of
the experimentally observed population redistribution into
states. Their fitting method can then
serve as a tool to estimate the electron temperature.

\item
 The plasma expansion itself is already a signature of the (time average) temperature as indicated by equations (\ref{u_self_sim}) and (\ref{expansion}). One promising way toward this measurement has been provided by \citealt{killian2003} who use Strontium ions. Indeed, the non alkali ions as the strontium one have visible resonance transitions and can then be detected by laser induced fluorescence. The Doppler shift of the resonance  indicates that the ion velocity, and so it's variation (the acceleration), reflects the electron temperature \citep{simien2004}.

For alkali atoms it is no more possible to use optical detection. We might think to use position sensitive detector to image the ionic cloud. A similar idea is to use the time of flight projection image of the ionic cloud. We have tried this technique. It is efficient and easy to handle only for really small ion number clouds to avoid a too strong Coulomb explosion during the ion transport toward the MCP. Indeed, even using formula (\ref{eq_spike}), which is based on a hypothetical pure spherical symmetry, it is hard to restore the initial ionic cloud shape from the projection image of the ions in the detector.

\item Another method  to study the plasma temperature is to monitor the electron evaporation. Experimentally, the easiest way is to add a small electric field to 
force slow evaporation of particles.
 Theoretically this is
the  well known Kramers problem of escape from a trap \citep{melnikov1991,hanggi1990}. 
The escape rate (see equation (\ref{loss_rate})), can be calculated for instance using a King distribution and is usually found to be close to $1/ (100 t_{\rm e})$ \citep{binney1987} where $t_e = 9 \sigma_v^3 / (16 \sqrt{\pi} G'^2 m_e^2 n_e^0 \ln \Lambda)  $ is the electron thermalisation time. This results assume 
a purely energetic truncated condition. But, in globular cluster dynamics it is well known that this hypothesis of purely energetic truncated condition is not well suited especially for anisotropic systems \citep{takahashi1997,kim1999}. The comparison between ultra-cold plasma experiments and the theories might help to understand how accurate the assumption of energy truncation is. This technique has been proposed by the NIST group using a constant small electric field to detect all the eject electrons. Some results are reported in \citealt{robicheaux2003} figure 3 but were not interpreted. The results are in fact compatible with the 
$1/ (100 t_{\rm e})$ evaporation rate on the order of $10^{11}\,$s$^{-1}$ decreasing with time due to the fact the density decrease much faster than the electron average velocity during the plasma expansion.

\item Similarly to our work, \citealt{roberts2004} used electric field pulse to extract the plasma temperature. To avoid any discussion concerning the potential shape they tried to affect electrons only in the asymptotic part (in $1/r$) of the trapping potential bu using only  very small electric fields.   
 Their nice experimental data are treated using a theory based on maxwellian electron distribution which we believed is not as appropriate as the Kramers-King's one, especially for the slightly bound electrons which have been removed by the electric pulse. 
 Furthermore,  even with a small electron fraction removed, $N_i - N_e$ can be affected and the electron extraction process is probably very complex.
 The main results concerns the electron cooling during the plasma expansion and the almost constant initial temperature $T_e \approx 50 $K independently of $T_e^\gamma$. This final results looks surprising to us because it seems to disagree with the Virial based equations (\ref{virial_res}) (\ref{temp_king}) or (\ref{temp_simpl}). Unfortunately, the published data are not sufficient to check the results with other electron distribution.

\end{itemize}

\section{Conclusion}

We have shown that
the dynamical behavior of the electrons in an ultracold plasma 
is similar to the one described by conventional models of stars clusters dynamics. 
The evolution of such a sample is dominated
by kinetics equations formally identical to the ones controlling the evolution of a stars cluster using only a new negative gravitational constant defined by equation (\ref{replacement}).
We have developed here some aspects of the analogy with globular star cluster dynamics. The influence of stellar encounters namely: relaxation, equipartition, mass segregation, escape, inelastic encounters (coalescence, dissipation of energy), binary formation by three body encounters, interaction with primordial binaries stars,... have exact partners in the plasma case.
There is similar law for the three body recombination for instance the Heggie's law: ''hard binaries get harder and soft binaries get softer'' is identical in both system. Thanks to the Virial theorem we could relate the plasma temperature with the ions and electrons numbers  (or central density) 
and with the initial kinetic energy given to electrons by the laser ionization.
We found that
the 
Fokker-Planck equation is exactly identical for electron in ultra cold plasma and for stars in globular cluster.
 The only modification occurs in the potential which,  in the plasma case, should contain an extra part due to the ions.
The ions are spectator but are needed to create a confining potential for the electrons.
The gaseous equations, coming from the first moments equations of the Fokker-Planck equation, confirm the early self similar evolution of the ionic cloud. This self similar evolution breakdown after few microsecond and we have analytically studied the ion shock wave occurring when the quasineutrality is violated. The electrons density evolution can be studied using similar fluid equations than the one used for clusters. The orbit average study of the Fokker-Planck equation indicates that the
 quasi-static solution for the phase space density electron distribution $f$ is a 
 Kramers-King's one $f\propto e^{{\cal E}/(K_B T_e^K)} -1$ where $T_e^K$ deals for King (or Kramers) temperature although strictly speaking a thermodynamics temperature is not defined for a nonequilibrium distribution. This is confirmed by comparison with Monte Carlo simulation previously published. Even when the three body collisions are taken into account this approximate distribution seems accurate.
 In fact several numerical codes developed for stellar system could be easily adapted to treat the ultracold plasma case.
  Finally, it is possible to experimentally  simulate the tidal escape from globular cluster under external galaxy gravitational attraction by
using small external field to extract electron trapped in an ionic cloud. The full process is beyond the scope of the article but might be tackled with the violent relaxation theory. We have only study here the link between the electrons and ions numbers, the plasma temperature and the charged particles potential depth through formula (\ref{temp_simpl}).

We hope that this article will stimulate links with the astrophysics community and we hope the analogy is also useful to develop more physical insight on the ultra-cold plasma physics behavior.
For instance as in the cluster case \cite{meylan1997} binary collisions might be a route for the formation of long-lived multiple systems as three-body stable systems or for the formation of giant molecules as ''trilobites'' \citep{greene2000} or macro-Rydberg ones in the plasma case \citep{farooqi2003}.    
The ultracold plasma experiments evolve rapidly, we have mentioned the  laser manipulation of the ions in the strontium case \citep{simien2004}. This might lead to control of the ion motions and to cool sample where crystallization might appear \citep{pohl2004}. 
 This manipulation of ions can be added with external field manipulation as RF electric field, magnetic gradient manipulation  and might be able to simulate a lot of events as tidal shock or tidal heating in cluster physics.
 Some pictures of the ionic cloud have already been done by \cite{simien2004} and this gives exactly the projected density profile in strong analogy with the main observed quantity in star cluster: the projected mass profile.  A charge particle detector (MCP) with position sensitivity might also give similar data.
Therefore, analogy with dynamics of globular clusters might stimulate new ideas in ultra-cold plasma physics.
One advantage of plasma system compared to globular cluster one results in its physical behavior very closely related to what can be simulated. But, the major advantage is the experimental capability of tuning different parameters in a
huge range.
In globular cluster, non evolutionary models or single mass models are often used but are far from reality, this is no more the case in ultra-cold plasma physics. We believe that a comparison between different models with very well controlled experiments will help to improve further development. 

\section*{Acknowledgments}

 We would like to thanks Gary Manon, Rainer Spurtzem and Haldan Cohn for helpful discussions.  

\bibliographystyle{astron}
\bibliography{bibli}

\begin{thebibliography}{}

\bibitem[\protect\astroncite{Berg-Sorensen}{1997}]{berg-sorensen1997}
Berg-Sorensen, K.: 1997,
\newblock {\em Phys. Rev. A} {\bf 55(2)}, 1281,
\newblock Erratum in PRA 56 3308 (1997)

\bibitem[\protect\astroncite{Bergeson and Spencer}{2003}]{bergeson2003}
Bergeson, S.~D. and Spencer, R.~L.: 2003,
\newblock {\em Phys. Rev. E} {\bf 67}, 126414

\bibitem[\protect\astroncite{Binney and Tremaine}{1987}]{binney1987}
Binney, J. and Tremaine, S.: 1987,
\newblock {\em Galactic Dynamics},
\newblock Princeton Series in Astrophysics

\bibitem[\protect\astroncite{Chavanis}{1998}]{chavanis1998}
Chavanis, P.~H.: 1998,
\newblock {\em Mon. Not. R. Astron. Soc.} {\bf 300}, 981

\bibitem[\protect\astroncite{Chavanis}{2002}]{chavanis2002}
Chavanis, P.~H.: 2002,
\newblock Springer, Berlin,
\newblock astro-ph/0212205

\bibitem[\protect\astroncite{Cohn}{1979}]{cohn1979}
Cohn, H.: 1979,
\newblock {\em Astrophys. J.} {\bf (234)}, 1036

\bibitem[\protect\astroncite{Cohn et~al.}{1989}]{cohn1989}
Cohn, H., Hut, P., and Wise, M.: 1989,
\newblock {\em Astrophys. J.} {\bf (342)}, 814

\bibitem[\protect\astroncite{Delcroix and Bers}{1994}]{delcroix1994}
Delcroix, J.~L. and Bers, A.: 1994,
\newblock {\em Physique des plasmas},
\newblock EDP Science

\bibitem[\protect\astroncite{Dorozhkina and Semenov}{1998}]{dorozhkina1998}
Dorozhkina, D.~S. and Semenov, V.~E.: 1998,
\newblock {\em Phys. Rev. Lett.} {\bf 81(13)}, 2691

\bibitem[\protect\astroncite{Doyle et~al.}{1989}]{doyle1989}
Doyle, J.~M., Sandberg, J.~C., Masuhara, N., Yu, I.~A., Kleppner, D., and
  Greytak, T.~J.: 1989,
\newblock {\em J. Opt. Soc. Am. B} {\bf 6(11)}, 2244

\bibitem[\protect\astroncite{Dubin and O'Neil}{1999}]{dubin1999}
Dubin, D. H.~E. and O'Neil, T.~M.: 1999,
\newblock {\em Rev. Mod. Phys.} {\bf 71}, 87

\bibitem[\protect\astroncite{Dutta et~al.}{2001}]{dutta2001}
Dutta, S.~K., Feldbaum, D., Walz-Flannigan, A., Guest, J.~R., and Raithel, G.:
  2001,
\newblock {\em Phys. Rev. Lett.} {\bf 86(18)}, 3993

\bibitem[\protect\astroncite{Farooqi et~al.}{2003}]{farooqi2003}
Farooqi, S.~M., Tong, D., Krishnan, S., Stanojevic, J., Zhang, Y.~P., Ensher,
  J.~R., Estrin, A.~S., Boisseau, C., Côté, R., Eyler, E.~E., and Gould, P.~L.:
  2003,
\newblock {\em Phys. Rev. Lett.} {\bf 91}, 183002

\bibitem[\protect\astroncite{Fregeau et~al.}{2004}]{fregeau2004}
Fregeau, J.~M., Cheug, P., Zwart, S. F.~P., and Rasio, F.~A.: 2004,
\newblock {\em Mon. Not. R. Astron. Soc.},
\newblock arXiv:astro-ph/0401004

\bibitem[\protect\astroncite{Gallagher et~al.}{2003}]{gallagher2003}
Gallagher, T.~F., Pillet, P., Robinson, M.~P., Laburthe-Tolra, B., and Noel,
  M.~W.: 2003,
\newblock {\em J. Opt. Soc. Am. B} {\bf 20(6)}, 1091

\bibitem[\protect\astroncite{Giersz and Spurzem}{2004}]{giersz2004}
Giersz, M. and Spurzem, R.: 2004,
\newblock {\em Mon. Not. R. Astron. Soc.},
\newblock arXiv:astro-ph/0301643

\bibitem[\protect\astroncite{Goodman and Hut}{1993}]{goodman1993}
Goodman, J. and Hut, P.: 1993,
\newblock {\em The Astrophysical Journal} {\bf 403}, 271

\bibitem[\protect\astroncite{Greene et~al.}{2000}]{greene2000}
Greene, C.~H., Dickinson, A.~S., and Sadeghpour, H.~R.: 2000,
\newblock {\em Phys. Rev. Lett.} {\bf 85}, 2458

\bibitem[\protect\astroncite{Hahn}{2000}]{hahn2000}
Hahn, Y.: 2000,
\newblock {\em J. Phys. B} {\bf 33}, L655

\bibitem[\protect\astroncite{Hahn}{2002}]{hahn2002}
Hahn, Y.: 2002,
\newblock {\em Phys. Lett. A} {\bf 293}, 266

\bibitem[\protect\astroncite{Heggie}{1975}]{heggie1975}
Heggie, D.~C.: 1975,
\newblock {\em Mon. Not. R. Astron. Soc.} {\bf 173}, 729

\bibitem[\protect\astroncite{Heggie et~al.}{1998}]{heggie1998}
Heggie, D.~C., Giersz, M., Spurzem, R., and Takahashi, K.: 1998,
\newblock {\em Highlights of Astronomy} {\bf 11(A)}, 591

\bibitem[\protect\astroncite{Heggie and Stevenson}{1988}]{heggie1988}
Heggie, D.~C. and Stevenson, D.: 1988,
\newblock {\em Mon. Not. R. Astron. Soc.} {\bf 230}, 223

\bibitem[\protect\astroncite{H\'enon}{1960}]{henon1960}
H\'enon, M.: 1960,
\newblock {\em Annales d'Astroph.} {\bf 23}, 668

\bibitem[\protect\astroncite{H\'enon}{1961}]{henon1961}
H\'enon, M.: 1961,
\newblock {\em Annales d'Astroph.} {\bf 24(5)}, 369

\bibitem[\protect\astroncite{Henyey et~al.}{1964}]{henyey1964}
Henyey, L.~G., Forbes, J.~E., and Gould, N.~L.: 1964,
\newblock {\em The Astrophysical Journal} {\bf 139}, 306

\bibitem[\protect\astroncite{HQ et~al.}{2001}]{hua2001}
HQ, H., JL, B., and L, W.: 2001,
\newblock {\em Chem. Phys.} {\bf 270}, 93

\bibitem[\protect\astroncite{Hut and Bahcall}{1983}]{hut1983}
Hut, P. and Bahcall, J.~N.: 1983,
\newblock {\em The Astrophysical Journal} {\bf 268}, 319

\bibitem[\protect\astroncite{Hut et~al.}{1992}]{hut1992}
Hut, P., McMillan, S., Goodman, J., Mateo, M., Phinney, E.~S., Pryor, C.,
  Richer, H.~B., Verbunt, F., and Weinberg, M.: 1992,
\newblock {\em 'PASP'} {\bf 104}, 981

\bibitem[\protect\astroncite{Inagaki and Lynden-Bell}{1990}]{inagaki1990}
Inagaki, S. and Lynden-Bell, D.: 1990,
\newblock {\em Mon. Not. R. astr. Soc.} {\bf 244}, 254

\bibitem[\protect\astroncite{Johnstone}{1993}]{johnstone1993}
Johnstone, D.: 1993,
\newblock {\em Astronomical Journal} {\bf 105(1)}, 155

\bibitem[\protect\astroncite{Kaplan et~al.}{2003}]{kaplan2003}
Kaplan, A.~E., Dubetsky, B.~Y., and Shkolnikov, P.~L.: 2003,
\newblock {\em Phys. Rev. Lett.} {\bf 91(14)}, 143401

\bibitem[\protect\astroncite{Killian et~al.}{2003}]{killian2003}
Killian, T.~C., Ashoka, V.~S., Gupta, P., Laha, S., Nagel, S.~B., Simien,
  C.~E., Kulin, S., Rolston, S.~L., and Bergeson, S.~D.: 2003,
\newblock {\em J. Phys. A: Math. Gen.} {\bf 36}, 6077

\bibitem[\protect\astroncite{Killian et~al.}{1999}]{kulin1999}
Killian, T.~C., Kulin, S., Bergeson, S.~D., Orozco, L.~A., Orzel, C., and
  Rolston, S.~L.: 1999,
\newblock {\em Phys. Rev. Lett.} {\bf 83(23)}, 4776

\bibitem[\protect\astroncite{Killian et~al.}{2001}]{killian2001}
Killian, T.~C., Lim, M.~J., Kulin, S., Dumke, R., Bergeson, S.~D., and Rolston,
  S.~L.: 2001,
\newblock {\em Phys. Rev. Lett.} {\bf 86(17)}, 3759

\bibitem[\protect\astroncite{Kim and Oh}{1999}]{kim1999}
Kim, Y.~K. and Oh, K.~S.: 1999,
\newblock {\em J. Kor. Astr. Soc.} {\bf 32}, 17

\bibitem[\protect\astroncite{King}{1965}]{king1965}
King, I.~R.: 1965,
\newblock {\em Astronomical Journal} {\bf (70)}, 376

\bibitem[\protect\astroncite{King}{1966}]{king1966}
King, I.~R.: 1966,
\newblock {\em Astronomical Journal} {\bf (71)}, 64

\bibitem[\protect\astroncite{Kovalev and Bychenkov}{2003}]{kovalev2003}
Kovalev, V.~F. and Bychenkov, V.~Y.: 2003,
\newblock {\em Phys. Rev. Lett.} {\bf 90(18)}, 185004

\bibitem[\protect\astroncite{Kulin et~al.}{2000}]{kulin2000}
Kulin, S., Killian, T.~C., Bergeson, S.~D., and Rolston, S.~L.: 2000,
\newblock {\em Phys. Rev. Lett.} {\bf 85(2)}, 318

\bibitem[\protect\astroncite{Kulsrud et~al.}{1973}]{kulsrud1973}
Kulsrud, R.~M., Sun, Y.-C., Winsor, N.~K., and Fallon, H.~A.: 1973,
\newblock {\em Phys. Rev. Lett.} {\bf 31(11)}, 690

\bibitem[\protect\astroncite{Kuzmin and O'Neil}{2002a}]{kuzmin2002b}
Kuzmin, S.~G. and O'Neil, T.~M.: 2002a,
\newblock {\em Phys. plasmas} {\bf 9(9)}, 3743

\bibitem[\protect\astroncite{Kuzmin and O'Neil}{2002b}]{kuzmin2002a}
Kuzmin, S.~G. and O'Neil, T.~M.: 2002b,
\newblock {\em Phys. Rev. Lett.} {\bf 88(6)}, 065003

\bibitem[\protect\astroncite{Larson}{1970}]{larson1970}
Larson, R.~B.: 1970,
\newblock {\em Mon. Not. R. Astron. Soc.} {\bf 147}, 323

\bibitem[\protect\astroncite{Li et~al.}{2004}]{li2004}
Li, W., Noel, M.~W., Robinson, M.~P., Tanner, P.~J., Gallagher, T.~F., and.
  B.~Laburthe~Tolra, D.~C., Vanhaecke, N., Vogt, T., Zahzam, N., Pillet, P.,
  and Tate, D.~A.: 2004,
\newblock {\em Phys. Rev. A},
\newblock submitted

\bibitem[\protect\astroncite{Louis}{1990}]{louis1990}
Louis, P.~D.: 1990,
\newblock {\em Mon. Not. R. Astron. Soc.} {\bf 244}, 478

\bibitem[\protect\astroncite{Louis and Spurzem}{1991}]{louis1991}
Louis, P.~D. and Spurzem, R.: 1991,
\newblock {\em Mon. Not. R. Astron. Soc.} {\bf 251}, 408

\bibitem[\protect\astroncite{Lyman~Spitzer}{1987}]{spitzer1987}
Lyman~Spitzer, J.: 1987,
\newblock {\em Dynamical evolution of globular clusters},
\newblock Princeton University Press, Princeton, New Jersey

\bibitem[\protect\astroncite{Lynden-Bell and Eggleton}{1980}]{lynden1980}
Lynden-Bell, D. and Eggleton, P.~P.: 1980,
\newblock {\em Mon. Not. R. Astron. Soc.} {\bf 191}, 483

\bibitem[\protect\astroncite{Majumdar et~al.}{1973}]{majumdar1971}
Majumdar, S.~K., Adhikari, D., and Lahiri, A.: 1973,
\newblock {\em Plasma Physics} {\bf 15}, 1259

\bibitem[\protect\astroncite{Mansbach and Keck}{1969}]{mansbach1969}
Mansbach, P. and Keck, J.: 1969,
\newblock {\em Phys. Rev.} {\bf 181}, 275

\bibitem[\protect\astroncite{Mazevet et~al.}{2002}]{mazevet2002}
Mazevet, S., Collins, L.~A., and Kress, J.~D.: 2002,
\newblock {\em Phys. Rev. Lett.} {\bf 88}, 055001

\bibitem[\protect\astroncite{Mel'nikov}{1991}]{melnikov1991}
Mel'nikov, V.~I.: 1991,
\newblock {\em Phys. Rep.} {\bf 209(1-2)}, 1

\bibitem[\protect\astroncite{Meylan and Heggie}{1997}]{meylan1997}
Meylan, G. and Heggie, D.~C.: 1997,
\newblock {\em The Astron. Astrophys. Rev.} {\bf 8}, 1

\bibitem[\protect\astroncite{Mitchner and Kruger}{1992}]{mitchner1992}
Mitchner, M. and Kruger, H.~C.: 1992,
\newblock {\em Partially ionized gases},
\newblock Department of Mechanical Engineering, Stanford University,
\newblock avalaible at http://navier/stanford.edu/PIG/PIGdefault.html

\bibitem[\protect\astroncite{Mora}{2003}]{mora2003}
Mora, P.: 2003,
\newblock {\em Phys. Rev. Lett.} {\bf 90(18)}, 185002

\bibitem[\protect\astroncite{Peter et~al.}{1990}]{hanggi1990}
Peter, H., Peter, T., and Michal, B.: 1990,
\newblock {\em Rev. Mod. Phys.} {\bf 62(2)}, 251

\bibitem[\protect\astroncite{Pinke et~al.}{1998}]{pinke1998}
Pinke, P. W.~H., Mosk, A., Weidem\"{u}ller, M., Reynolds, M.~W., Hijmans,
  T.~W., and Walraven, J. T.~M.: 1998,
\newblock {\em Phys. Rev. A} {\bf 57(6)}, 4747

\bibitem[\protect\astroncite{Pohl et~al.}{2004a}]{pohl2004d}
Pohl, T., Pattard, T., and Rost, J.: 2004a,
\newblock arXiv:physics/0410016

\bibitem[\protect\astroncite{Pohl et~al.}{2004b}]{pohl2004c}
Pohl, T., Pattard, T., and Rost, J.: 2004b,
\newblock arXiv:physics/0405125

\bibitem[\protect\astroncite{Pohl et~al.}{2004c}]{pohl2004}
Pohl, T., Pattard, T., and Rost, J.: 2004c,
\newblock arXiv:physics/0402010, accepted to J. Phys. B

\bibitem[\protect\astroncite{Pohl et~al.}{2003}]{pohl2003}
Pohl, T., Pattard, T., and Rost, J.~M.: 2003,
\newblock {\em Phys. Rev. A} {\bf 68}, 010703

\bibitem[\protect\astroncite{Pohl et~al.}{2004d}]{pohl2004b}
Pohl, T., Pattard, T., and Rost, J.~M.: 2004d,
\newblock arXiv:physics/0311131, accepted to PRL

\bibitem[\protect\astroncite{Prata}{1971}]{prata1971}
Prata, S.~W.: 1971,
\newblock {\em Astronomical Journal} {\bf (76)}, 1029

\bibitem[\protect\astroncite{Roberts et~al.}{2004}]{roberts2004}
Roberts, J.~L., Fertig, C.~F., Lim, M.~L., and Rolston, S.~L.: 2004,
\newblock {\em Phys. Rev. Lett.} {\bf 92}, 253003

\bibitem[\protect\astroncite{Robicheaux and Hanson}{2003}]{robicheaux2003}
Robicheaux, F. and Hanson, D.: 2003,
\newblock {\em Phys. plasmas} {\bf 10(6)}, 2217

\bibitem[\protect\astroncite{Robicheaux and Hanson}{2002}]{robicheaux2002}
Robicheaux, F. and Hanson, J.~D.: 2002,
\newblock {\em Phys. Rev. Lett.} {\bf 88(5)}, 055002

\bibitem[\protect\astroncite{Robinson et~al.}{2000}]{robinson2000}
Robinson, M.~P., Tolra, B.~L., Noel, M.~W., Gallagher, T.~F., and Pillet, P.:
  2000,
\newblock {\em Phys. Rev. Lett.} {\bf 85(21)}, 4466

\bibitem[\protect\astroncite{Rosenbluth et~al.}{1957}]{rosenbluth1957}
Rosenbluth, M.~N., MacDonald, W.~M., and Judd, D.~L.: 1957,
\newblock {\em Phys. Rev.} 107(1)

\bibitem[\protect\astroncite{Sigurdsson and Phinney}{1998}]{sigurdson1993}
Sigurdsson, S. and Phinney, E.~S.: 1998,
\newblock {\em Astrophys. J.} {\bf 500}, 130

\bibitem[\protect\astroncite{Simien et~al.}{2004}]{simien2004}
Simien, C., Chen, Y., Gupta, P., Laha, S., Martinez, Y., Mickelson, P., Nagel,
  S., and Killian, T.: 2004,
\newblock {\em Phys. Rev. Lett.} {\bf 92(14)}, 143001

\bibitem[\protect\astroncite{Spitzer and Shapiro}{1972}]{spitzer1972}
Spitzer, L. and Shapiro, S.~L.: 1972,
\newblock {\em Astrophys. J.} {\bf 173}, 529

\bibitem[\protect\astroncite{Stevefelt et~al.}{1975}]{stevefelt1975}
Stevefelt, J., Boulmer, J., and Delpech, J.-F.: 1975,
\newblock {\em Phys. Rev. A} {\bf 12}, 1246

\bibitem[\protect\astroncite{Surkov et~al.}{1996}]{surkov1996}
Surkov, E.~L., Walraven, J. T.~M., and Shlyapnikov, G.~V.: 1996,
\newblock {\em Phys. Rev. A} {\bf 53(5)}, 3403

\bibitem[\protect\astroncite{Takahashi}{1993}]{takahashi1993}
Takahashi, K.: 1993,
\newblock {\em PASJ: Publications of the Astronomical Society of Japan} {\bf
  2(45)}, 233

\bibitem[\protect\astroncite{Takahashi et~al.}{1997}]{takahashi1997}
Takahashi, K., Lee, H.~M., and Inagaki, S.: 1997,
\newblock {\em Mon. Not. R. Astron. Soc.} {\bf 292}, 331

\bibitem[\protect\astroncite{Takahashi and Zwart}{1998}]{takahashi1998}
Takahashi, K. and Zwart, S. F.~P.: 1998,
\newblock {\em Astrophys. J.} {\bf 502}, L49

\bibitem[\protect\astroncite{Tkachev and Yakovlenko}{2001a}]{tkachev2001a}
Tkachev, A.~N. and Yakovlenko, S.~I.: 2001a,
\newblock {\em Quantum Electron.} {\bf 73(2)}, 66

\bibitem[\protect\astroncite{Tkachev and Yakovlenko}{2001b}]{tkachev2001}
Tkachev, A.~N. and Yakovlenko, S.~I.: 2001b,
\newblock {\em Quantum Electron.} {\bf 31(12)}, 1084

\bibitem[\protect\astroncite{Tol et~al.}{2004}]{tol2004}
Tol, P. J.~J., Hogervorst, W., and Vassen, W.: 2004,
\newblock {\em Phys. Rev. A} {\bf 70}, 013404

\bibitem[\protect\astroncite{Vanhaecke et~al.}{2004}]{vanhaecke2004}
Vanhaecke, N., Comparat, D., Tate, D.~A., and Pillet, P.: 2004,
\newblock arXiv:quant-ph/0401046, PRA accepted

\bibitem[\protect\astroncite{Vriens and Smeets}{1980}]{vriens1980}
Vriens, L. and Smeets, A. H.~M.: 1980,
\newblock {\em Phys. Rev. A} {\bf 22}, 940

\bibitem[\protect\astroncite{Walz-Flannigan et~al.}{2004}]{walz2004}
Walz-Flannigan, A., Guest, J.~R., Choi, J.-H., and Raithel, G.: 2004,
\newblock {\em Phys. Rev. A} {\bf 69}, 063405

\bibitem[\protect\astroncite{Wiechen and Ziegler}{1994}]{wiechen1994}
Wiechen, H. and Ziegler, H.~J.: 1994,
\newblock {\em J. Plasma Physics} {\bf 51(2)}, 341

\bibitem[\protect\astroncite{Wiyanto et~al.}{1985}]{wiyanto1985}
Wiyanto, P., Kato, S., and Inagaki, S.: 1985,
\newblock {\em PASJ: Publications of the Astronomical Society of Japan} {\bf
  37}, 715

\bibitem[\protect\astroncite{Ziegler and Wiechen}{1989}]{ziegler1989}
Ziegler, H.~J. and Wiechen, H.: 1989,
\newblock {\em Mon. Not. R. Astron. Soc.} {\bf 238(1)}, 1261

\end{thebibliography}

\addcontentsline{toc}{chapter}{Bibliographie}

\label{lastpage}

\end{document}